\pdfoutput=0
\documentclass[11pt]{article}
\usepackage{epsfig}
\usepackage{amsfonts}
\usepackage{amsmath}
\usepackage{bm,bbm}
\usepackage{cite}
\usepackage{cancel}
\usepackage[hidelinks]{hyperref}
\allowdisplaybreaks[1]

\newcommand{\la}[1]{\label{#1}}
\newcommand{\be}{\begin{equation}}
\newcommand{\ee}{\end{equation}}
\newcommand{\ba}{\begin{eqnarray}}
\newcommand{\ea}{\end{eqnarray}}
\newcommand{\rmi}[1]{{\mbox{\scriptsize #1}}}
\newcommand{\fig}{Fig.~}

\newcommand{\eq}{Eq.~}
\newcommand{\eqs}{Eqs.~}
\newcommand{\se}{Sec.~}

\newcommand{\nr}[1]{(\ref{#1})}

\newcommand{\nn}{\nonumber \\}
\newcommand{\fr}[2]{{\frac{#1}{#2}\,}}

\renewcommand{\vec}[1]{{\bf #1}}

\newcommand{\tfr}[2]{{\textstyle \frac{#1}{#2}\,}}

\renewcommand{\eq}{eq.~}
\renewcommand{\eqs}{eqs.~}
\renewcommand{\se}{sec.~}

\renewcommand{\fig}{fig.~}

\newcommand{\B}{\rmii{$B$}}
\newcommand{\E}{\rmii{$E$}}

\newcommand{\sH}{\rmii{$H$}}

\newcommand{\T}{\rmii{$T$}}
\newcommand{\R}{\rmii{$R$}}

\newcommand{\mpl}{m_\rmi{pl}}

\newcommand{\Tc}{T_{\rm c}}

\newcommand{\rmO}{{\mathcal{O}}}

\def\lsi{\raise0.3ex\hbox{$<$\kern-0.75em\raise-1.1ex\hbox{$\sim$}}}
\def\gsi{\raise0.3ex\hbox{$>$\kern-0.75em\raise-1.1ex\hbox{$\sim$}}}

\newcommand{\gsim}{\mathop{\gsi}}

\newcommand{\nB}{n_\rmii{B}} 
\newcommand{\rmii}[1]{{\mbox{\tiny\rm{#1}}}}

\newcommand{\im}{\mathop{\mbox{Im}}}

\newcommand{\Tint}[1]{{\hbox{$\sum$}\!\!\!\!\!\!\!\int\,}_{\!\!\!\!\raise-0.9ex\hbox{$\scriptstyle{#1}$}}}
\newcommand{\Tinti}[1]{{{\Sigma}\!\!\!\!\raise0.3ex\hbox{$\int$}_\rmii{${#1}$}}}

\newcommand{\bi}{\begin{itemize}}
\newcommand{\ei}{\end{itemize}}
\newcommand{\hide}[1]{ }
\newcommand{\blind}[1]{\fbox{$ ? $}} 

\newcommand{\deltabar}{\raise-0.02em\hbox{$\bar{}$}\hspace*{-0.8mm}{\delta}}

\newcommand{\mH}{m_\rmii{$H$}} 
\newcommand{\mW}{m_\rmii{$W$}} 

\newcommand{\Q}{\mathcal{Q}}
\renewcommand{\O}{\mathcal{O}}
\newcommand{\K}{\mathcal{K}}
\newcommand{\X}{\mathcal{X}}
\newcommand{\Y}{\mathcal{Y}}
\newcommand{\iT}{\rmii{$T$}}
\newcommand{\aS}{h} 
\newcommand{\mS}{m_\aS}

\newcommand{\vw}{v_\rmi{w}}
\newcommand{\lwall}{\ell^{ }_\rmi{wall}}
\newcommand{\frict}{\Upsilon} 
\newcommand{\arxiv}[1]{[\href{https://arxiv.org/abs/#1}{#1}].}

\makeatletter \@addtoreset{equation}{section} \makeatother
\renewcommand{\theequation}{\arabic{section}.\arabic{equation}}
\makeatletter
\renewcommand\section{\@startsection {section}{1}{\z@}%
                                   {-5.5ex \@plus -1ex \@minus -.2ex}
                                   {2.3ex \@plus.2ex}%
                                   {\normalfont\large\bfseries}}
\renewcommand\subsection{\@startsection{subsection}{2}{\z@}%
                                     {-3.25ex\@plus -1ex \@minus -.2ex}%
                                     {1.5ex \@plus .2ex}%
                                     {\normalfont\normalsize\bfseries}}
\renewcommand\thesection {\@arabic\c@section}
\renewcommand\thesubsection   {\thesection.\@arabic\c@subsection}
\renewcommand{\@seccntformat}[1]{%
\csname the#1\endcsname.\hspace{1.0em}}
\makeatother

\begin{document}

\flushbottom

\begin{titlepage}

\begin{flushright}
September 2025
\end{flushright}
\begin{centering}
\vfill

{\Large{\bf
  Entropy production 
  at electroweak bubble walls \\[3mm]
  from scalar field fluctuations
}} 

\vspace{0.8cm}

M.~Eriksson, 
M.~Laine 

\vspace{0.8cm}

{\em
AEC, 
Institute for Theoretical Physics, 
University of Bern, \\ 
Sidlerstrasse 5, CH-3012 Bern, Switzerland \\}

\vspace*{0.8cm}

\mbox{\bf Abstract}
 
\end{centering}

\vspace*{0.3cm}
 
\noindent
The real-time dynamics of 
an electroweak phase transition involves 
large time and distance scales, the domain of hydrodynamics.
However, the matching conditions of ideal hydrodynamics across 
a bubble wall do not fix the fluid profile completely, with the remaining
degree of freedom parametrizable through entropy production. 
Within a framework of Langevin dynamics, viewed as
an effective description valid between the hydrodynamic 
($k \sim g^4_{ } T/\pi^3_{ }$)
and soft momentum scales ($k \sim gT$), 
we determine the entropy production originating 
from scalar field fluctuations. 
The entropy discontinuity is shown to remain non-vanishing
when the friction coefficient
is sent to zero, in apparent violation of  
the ``local thermal equilibrium'' (LTE) framework. 
To confirm the finding, we identify its origin 
within Boltzmann equations, as being part of 
the $1\to 1$ force associated with the ``ballistic'' regime. 
The result implies that LTE-based upper bounds 
on the wall velocity cannot be saturated. 

\vfill


\end{titlepage}

\tableofcontents

%
\section{Introduction}
\la{se:intro}

Given its gravitational 
wave signatures~\cite{lisa_pt}, 
as well as its possible role in generating 
a baryon asymmetry~\cite{rev_bb}, the phase transition
associated with the setting in of the Higgs mechanism
has been intensively studied in the past decades~\cite{rev_mbh}. 
In the Standard Model, the phase transition is 
a smooth crossover~\cite{endpoint}, but many extended 
Higgs sectors lead to a first-order transition. 
Generally speaking, it is first-order transitions that
are cosmologically the most interesting ones, as they
may lead to large deviations from thermal equilibrium, 
and could subsequently leave over cosmological remnants.

Among the simplest possibilities for producing a strong 
first-order phase transition is to envisage that it takes
place in two steps~\cite{twostep1}. In the first step, 
a non-Standard Model scalar field takes an expectation value. 
In the second step, that expectation value goes away, and 
the Standard Model Higgs phenomenon sets it. The latter 
transition interpolates between two ``broken'' phases, 
separated by a barrier even at the tree level. In fact
the barrier can be so large that the system gets stuck
in the unphysical ground state~\cite{twostep2,twostep3}.
If there is sufficient freedom to tune the couplings, 
however, the transition may proceed, thereby possibly
manifesting an interesting scenario~\cite{twostep4,twostep5}.

The real-time dynamics of first-order phase transitions
represents a complicated topic.
Normally, the transition is envisaged to proceed 
through the nucleation of bubbles of the low-temperature
phase~\cite{landau5,eikr}; 
their subsequent growth and collisions~\cite{landau6,ikkl}; 
a stage with sound waves
(cf.,\ e.g.,\ ref.~\cite{simu} and references therein)
and turbulence
(cf.,\ e.g.,\ ref.~\cite{turb} and references therein); 
and a final relaxation
into the new ground state, as the Hubble expansion gradually
dilutes away the latent heat released in the transition~\cite{vw299}. 

While investigating the complete dynamics is challenging, 
given the multitude of length and time scales involved, 
it should be simpler to address a subpart of the dynamics, 
namely the stationary growth of bubbles of the low-temperature 
phase. In particular, a frequently stated goal is to determine the
wall velocity, $\vw^{ }$, at which the phase 
boundary propagates for a long while, between nucleations 
and collisions.

However, the wall velocity
is {\em not} an unambiguous property of the theory,  
unlike the critical temperature ($\Tc^{ }$), the latent heat ($L$), 
and the surface tension ($\sigma$). 
Rather, $v^{ }_{\rm w}$ depends on the temperature 
at which nucleations happen, and this in turn depends
on how the system is driven through the transition, notably
on the Hubble rate.\footnote{%
 This means that $v_\rmii{w}^{ }$ 
 depends on the overall energy density, 
 which is not a parameter of the electroweak theory
 (it can be affected, e.g.,\ by otherwise decoupled dark sectors). 
 } 
Also, the nucleation temperature itself 
depends implicitly and non-locally 
on the wall velocity, in that faster
growing bubbles fill the universe faster, so that
fewer bubbles suffice to complete the process. 
Furthermore, the growing bubbles have a non-trivial
hydrodynamic profile, consisting not only of the 
phase transition front but also of a shock wave 
preceding it (for ``subsonic deflagrations''~\cite{defl}) or a rarefaction
wave following it (for ``detonations''~\cite{deto}), or of 
a special class with both a shock and a rarefaction wave
(for ``supersonic deflagrations''~\cite{hks}). 

The global nature of the hydrodynamic profile,  
even if a text-book topic~\cite{landau6}, is easy to forget. 
Many microscopic attempts at computing
the wall velocity start from the intuitive picture
of balancing a driving pressure against a friction force, 
the latter being influenced by transition radiation  
(for recent work and references see, 
e.g.,\ refs.~\cite{vw2,vw3,vw45,vw5,vw8,vw8aa,%
 vw8a,vw8b,vw8c,vw8d,vw8e,vw8f,vw8g}). 
However, this picture is physically oversimplified.
In a deflagration bubble, the pressure is actually
larger in front of the wall (the medium having been
heated up by a shock front), yet the bubble keeps on growing, 
dissipating the latent heat released into
the shocked region that has not undergone any transition yet.  

The concept of a friction force is confusing also in that
steady-state profiles exist even without any physical friction
(see, e.g.,\ refs.~\cite{vw30,vw31,vw32,vw33,vw34,vw36,vw37,vw37a,vw38,vw39,%
vw41,vw42,vw43}
and references therein).
This is referred to as the approximation
of ``local thermal equilibrium''. 
The frictionless
approximation likely sets an upper bound on 
the wall velocity for a given nucleation temperature.
 
The purpose of the present paper is to interpolate, 
on the conceptual level, between microscopic studies of particle scatterings
in the wall frame, and macroscopic hydrodynamical investigations.
The missing ingredient is physics taking place at an intermediate
scale, which adjusts the temperature and flow velocity to their
proper hydrodynamic values. This equilibration dynamics can be 
described by a framework similar to 
fluctuating hydrodynamics~\cite{landau9}, 
but extended by a scalar field~\cite{hydro}. 
Fluctuating hydrodynamics implements the 
fluctuation-dissipation theorem, meaning that dissipative 
transport coefficients need to be accompanied by the 
corresponding stochastic noise terms. 
For a scalar field, this can be achieved with a Langevin equation.  
We were inspired by a classic
investigation of the shear viscosity ($\eta$)~\cite{shear}, 
which found that fluctuations
add a term $\sim T^6_{ }/\eta$ to a non-fluctuating 
bare value $\sim \eta$, implying that the overall viscosity
cannot be arbitrarily small.
We remark that scalar fluctuations affecting 
bubble walls is an old topic~\cite{hk}, recently
also studied in refs.~\cite{ae,ae2}, 
and that similar physics also plays a role in the context 
of reheating after inflation~\cite{alica}. 

Our presentation is organized as follows. 
After formulating the set of fluctuating equations
to be solved, in \se\ref{se:setup}, we describe 
their solution at next-to-leading order
in fluctuations, in \se\ref{se:algorithm}.
Our conclusions are collected in \se\ref{se:concl}. 
Appendix~A contains an intuitive derivation of the
stochastic force that appears in the Langevin equation.

\section{Setup}
\la{se:setup}

\subsection{Outline}

At the electroweak epoch, $T\sim 100$~GeV, the Hubble radius is 
$\ell^{ }_\sH \equiv H^{-1}_{ } \sim \mpl^{ } / T^2 \sim 10^{17}_{ } / T$.
For typical first-order phase 
transitions in extensions of the Standard Model, 
bubbles of the low-temperature phase nucleate at distances well within
the horizon,  
$\ell^{ }_\B  \sim 10^{-2...-6}_{ }  \ell^{ }_{\sH} 
\sim 10^{11 ... 15 }_{ } / T$, but much above 
the mean free path of 
particle scatterings, $\lambda \sim 1 / (\alpha^2 T)$, where 
$\alpha = g^2 / (4\pi) \sim 1/30$ is the weak gauge coupling. 
Therefore, the macroscopic theory of ideal hydrodynamics should be 
suitable for describing the bubble dynamics~\cite{landau6}. 
In classical nucleation 
theory the initial bubble size can be expressed in terms of
the latent heat and surface tension~\cite{landau5}, and is 
then also expected to be macroscopic, even if perhaps only
marginally so, with values $R^{ }_{\B,\rmi{initial}} \gsim 10^2_{ } / T$
found in model computations. 
Subsequently, the bubbles grow until they collide with each other, 
with $R^{ }_{\B,\rmi{final}} \sim \ell^{ }_\B$. 

In order to describe a phase transition, the corresponding order parameters
may be included as hydrodynamic degrees of freedom. In the following, 
we denote the scalar 
order parameter by $h$ (e.g., for the Standard Model Higgs field). 
The other dynamic variables are the temperature $T$ and fluid flow velocity
$\vec{v}$, the latter of which appears in the covariant combination
$u \equiv \gamma(1,\vec{v})$, where 
$\gamma \equiv 1/\sqrt{1-\vec{v}^2}$
is the Lorentz factor.

Like all effective descriptions, hydrodynamics is an expansion in gradients, 
and therefore applicable to slowly varying 
long-wavelength degrees of freedom. In the 
description of phase transitions, this leads to an apparent paradox. 
Namely, the transition front in which $h$ varies is narrow, 
$\lwall \sim 1 / m^{ }_h \ll 1/(\alpha^2 T)$
(cf.\ \fig\ref{fig:profile}). 
At the same time, 
$T$ and $v$ adjust 
themselves to their proper values at distance scales
$\sim 1/(\alpha^2 T) $.
Within hydrodynamics proper, this
paradox is solved by treating phase transition fronts as discontinuities, 
across which matching conditions are imposed
(cf.\ \se\ref{ss:macro}). 
However, the value of 
one of the discontinuities is left unspecified by ideal hydrodynamics,  
and for it we need 
a more microscopic description which is able to resolve
the phase transition front
(cf.\ \se\ref{ss:inter}).

\subsection{Macroscopic hydrodynamic description}
\la{ss:macro}

\begin{figure}[t]

\hspace*{-0.1cm}
\centerline{%
 \epsfysize=7.3cm\epsfbox{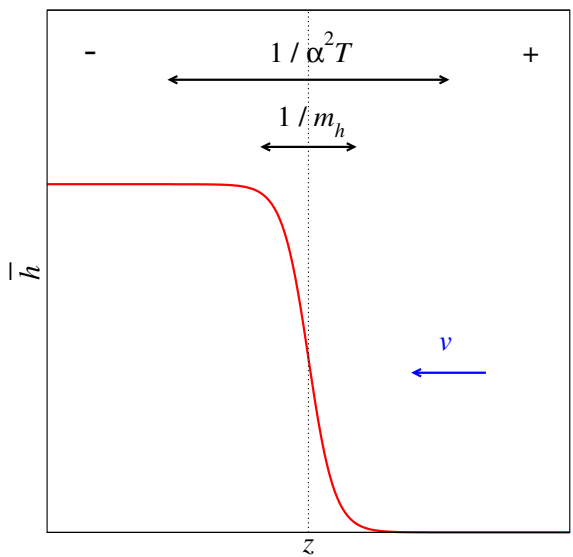}%
 \hspace{0.4cm}%
}

\caption[a]{\small
  Illustration of the background solution, 
  $\bar h(z)$,  in a planar wall rest frame. 
  With respect to the wall, the plasma moves to the left; 
  with respect to the plasma, the wall moves to the right.
  The wall profile is narrow compared with 
  hydrodynamic scales (cf.\ \eq\nr{hierarchy2}). We treat
  the $\gamma$-factor associated with the velocity 
  as being of $\rmO(1)$, however it could be 
  numerically largish. 
  }

\la{fig:profile}
\end{figure}

At large time and distance scales, above the mean free path of 
particle scatterings, energy and momentum are the only conserved 
quantities of our system, and the dynamics can be fixed
by postulating the form of the energy-momentum tensor. To leading
non-trivial order in gradients, we write 
\be
 T^{\mu\nu}_\rmi{ideal}
   \; \equiv \; 
   \overbrace{
   h^{,\mu}_{ }h^{,\nu}_{ }
 - \frac{ g^{\mu\nu}_{ } }{2}
     \,\Bigl(\,  
            h^{ }_{,\alpha} h^{,\alpha}_{ } 
     \,\Bigr) 
 }^{\equiv \; T^{\mu\nu}_{(h)} }
 \; + \;
 \overbrace{  
 ( e + p ) \,  u^{\mu}_{ }u^{\nu}_{ }+ p \, g^{\mu\nu}_{ }
 }^{\equiv \; T^{\mu\nu}_{(r)} }
 \;, \la{Tmunu_mixed}
\ee
where the scalar field $h$ supplements the temperature
and average flow velocity ($T$, $u^\mu_{ }$) as dynamical degrees
of freedom, 
$h^{,\mu}_{ } \equiv \partial^{\mu}_{ } h$, 
and the metric 
convention ($-$$+$$+$$+$) is assumed.
The pressure~$p$ and energy density~$e$ can be written as 
\be
 p \; \equiv \; p^{ }_r - V(h,T)
 \;, \quad
 e \; = \; T \partial^{ }_\T p - p 
 \; = \; e^{ }_r  + V - T V^{ }_{,\T}
 \;, \la{p_mixed}
\ee
where the contribution of ``radiation'' ($p^{ }_r$, $e^{ }_r$)
only depends on $T$, and we have defined partial derivatives as 
$
 (...)^{ }_{,x} \equiv \partial^{ }_x (...)
$.
The equations of motion take the form 
\be
 T^{\mu\nu}_{\rmi{ideal},\mu} = 0
 \;. \la{Tmunu_conserved}
\ee
We remark that there is not enough 
information in these equations to fix all the dynamical variables
(see below).

Let us now assume the presence of a bubble growing with constant velocity. 
We choose a patch of the phase transition front and boost to its rest frame,
in which the wall looks locally planar. The direction orthogonal to the 
surface is denoted by $z$; the situation is illustrated in 
\fig\ref{fig:profile}. In this frame the medium flows to the left, 
towards the stationary wall, 
so we write the four-velocity as 
\be
 u^0_{ } = \gamma \;, \quad
 u^z_{ } = - \gamma v \;, \quad
 v \equiv |\vec{v}|
 \;. \la{velocity}
\ee
In a static situation, \eq\nr{Tmunu_conserved} takes the form
\be
 \partial^{ }_z T^{zt}_\rmi{ideal} = 0
 \;, \quad
 \partial^{ }_z T^{zz}_\rmi{ideal} = 0
 \;, 
\ee
which can be integrated into the matching conditions 
\be
 T^{zt}_\rmi{ideal} \bigr|^{-}_{ } = 
 T^{zt}_\rmi{ideal} \bigr|^{+}_{ } 
 \;, \quad
 T^{zz}_\rmi{ideal} \bigr|^{-}_{ } = 
 T^{zz}_\rmi{ideal} \bigr|^{+}_{ } 
 \;, \la{bcs}
\ee
where ``$|^{\pm}_{ }$'' indicates that the variables are evaluated
at positions $z^{\pm}_{ }$ which are far away from the wall, 
$|z^{\pm}_{ }| \gg 1 / m^{ }_h $.
Inserting \eq\nr{Tmunu_mixed} and denoting $w\equiv e + p$, we thus 
find~\cite{landau6}
\ba
 \bigl( w \gamma^2_{ } v\bigr)^{-}_{ } 
 &
 \underset{\rmii{\nr{bcs}}}{
 \overset{\rmii{\nr{Tmunu_mixed}}}{=}} 
 & 
 \bigl( w \gamma^2_{ } v\bigr)^{+}_{ } 
 \; \equiv \; j 
 \; > \; 0
 \;, \la{disc1} \\ 
 \bigl( w \gamma^2_{ } v\bigr)^{-}_{ } v^{-}_{ } + p^{-}_{ } 
 & 
 \underset{\rmii{\nr{bcs}}}{
 \overset{\rmii{\nr{Tmunu_mixed}}}{=}} 
 & 
 \bigl( w \gamma^2_{ } v\bigr)^{+}_{ } v^{+}_{ } + p^{+}_{ }
 \;, \la{disc2}
\ea
out of which follows the important relation
\be
 \frac{p^+_{ } - p^-_{ }}{v^+_{ } - v^-_{ }}
 \; =  \;
 - j
 \;  < \; 
 0 
 \;.
\ee
In particular, considering a 
subsonic deflagration solution
in the global medium rest frame, 
the medium is 
moving in the radial direction in front of the wall, having been
kicked in motion by a shock front, but is at rest behind the wall, 
as required by spherical symmetry. 
Boosting to the wall frame this transforms to $v^+_{ } < v^-_{ }$, 
implying $p^+_{ } > p^-_{ }$. So, counter-intuitively,  
the pressure is larger in front of the wall. 

Let us now count the variables and constraints. 
Assuming that far enough from the wall, 
$h$ settles to the minimum of the effective potential, so that its
value is fixed by the temperature and is not independently dynamical, 
the matching conditions
relate four quantities, $T^{-}_{ }$, $T^{+}_{ }$, $v^-_{ }$ and $v^+_{ }$.
In a subsonic deflagration, $v^-_{ } = \vw^{ }$, while in a detonation,
$v^+_{ } = \vw^{ }$. 
The wall velocity $\vw^{ }$ is the main quantity that we want to determine.  
There is one variable which we may assume to be known as a boundary 
condition, namely the nucleation temperature, which would equal
$T^+_{ }$ in a detonation. In a deflagration, the nucleation 
temperature gets modified by a shock front, so it does not equal
$T^+_{ }$, but we may say that the nucleation temperature 
indirectly fixes $T^+_{ }$. 
In any case, three unknowns remain
($T^{-}_{ }$, $v^-_{ }$ and $v^+_{ }$), and there are only two matching
conditions, so one quantity remains undetermined. 

A physically meaningful way to fix the missing information is 
to determine the entropy released in the phase transition. The
entropy current reads $s u^\mu_{ }$, where $s = p^{ }_{,\iT}$ is the 
entropy density in the medium rest frame. Since the energy density
is the Legendre transform of the pressure, 
$e = T p^{ }_{,\iT} -p $, 
it follows that $w = e + p = Ts$.

In general, we expect
the local entropy production to be non-negative, i.e.\ 
$
 (s u^\mu_{ })^{ }_{,\mu} \ge 0
$.
In the frame of \fig\ref{fig:profile},  
this can be integrated to the total entropy change
\be
 \Delta^{ }_{s u^z_{ }} \; \equiv \; 
 \int_{z^{ }_-}^{z^{ }_+} \! {\rm d}z \, 
 (s u^z_{ })^{ }_{,z} 
 \quad 
 \underset{u^z_{ }\;=\;-\gamma v}{
 \overset{\rmii{\nr{velocity}}}{\Rightarrow}} 
 \quad 
 \bigl( s \gamma v \bigr)^-_{ } = 
 \bigl( s \gamma v \bigr)^+_{ } + \Delta^{ }_{s u^z_{ }}
 \;, 
 \quad
 \Delta^{ }_{s u^z_{ }} \ge 0 
 \;. \la{Delta_s_def}
\ee 
Determining $\Delta^{ }_{s u^z_{ }}$ requires microscopic information beyond
mere ideal hydrodynamics. We note that with \eq\nr{disc1}, and 
inserting $w = Ts$, \eq\nr{Delta_s_def} can be rephrased as 
\be
 \biggl( \frac{1}{T\gamma} \biggr)^{-}_{ }
 - 
 \biggl( \frac{1}{T\gamma} \biggr)^{+}_{ }
 \; 
 = 
 \; 
 \frac{\Delta^{ }_{s u^z_{ }}}{j}
 \; 
 \ge
 \; 
 0 
 \;. \la{T_disc}
\ee
This shows that~$T$ and~$v$ can be discontinuous even if
$\Delta^{ }_{s u^z_{ }} = 0$, for 
$
 T^{-}_{ }/ T^{+}_{ } 
 =
 \gamma^{+}_{ }/\gamma^{-}_{ }
$. 

\subsection{Intermediate Langevin description}
\la{ss:inter}

The definition of a microscopic description is not unique. 
In principle, we could start from the full quantum theory, 
but in an interacting thermal system this requires resummations
(such as the HTL and LPM resummations). 
It may be more economic to start from an effective theory, 
which already implements some resummations. However, in order to 
establish which effective description is the relevant one, we need
to specify the scale 
hierarchies present.

Around the electroweak phase transition, the thermal Higgs mass 
squared takes the form 
\be
 \mS^2 \sim -\frac{m^2_{\aS,0}}{2} + c \, g^2 T^2 
 \;, \la{mh}
\ee
where $m^{ }_{\aS,0}$ is the vacuum Higgs mass, 
and $c$ is a parameter of $\rmO(1)$~\cite{meg}. 
Close to the phase transition, 
the two terms in \eq\nr{mh} more or less cancel each other, 
but not completely. A rather trivial reason for 
the non-cancellation are  
higher-order corrections of $\rmO(g^3T^2/\pi)$~\cite{pba}, 
and the fact that
the dynamics becomes 
non-perturbative if $m_h^2 \sim \rmO(g^2T/\pi)^2$~\cite{linde}.
But it can also be said that  
in between the stable phases the curvature of the potential is negative, 
even if it were positive in the stable phases
on both sides of the transition. 
Such a ``tachyonic'' mode should conceivably be viewed 
as an IR degree of freedom. To formally justify such a picture, 
we postulate the hierarchy~\cite{gv2,lm,gt} 
\be
 \underbrace{ H }_{\rm Hubble~rate}
 \; \ll \; 
 \underbrace{ \alpha^2_{ } T }_{\rm hydrodynamics}
 \; \ll \; 
 \underbrace{ \alpha T }_{\rm Linde~scale}
 \; \ll \quad 
 m^{ }_{\aS}
 \quad \ll \; 
 \underbrace{ gT }_{\rm kinetic~theory}
 \; \ll \; 
 \underbrace{ \pi T }_{\rm full~theory}
 \;. \la{hierarchy2}
\ee
This implies that $h$ is a long-distance mode
from the kinetic theory point of view, 
so that Boltzmann equations~\cite{vw2,vw3,vw45,vw5,vw8,vw8aa,%
 vw8a,vw8b,vw8c,vw8d,vw8e,vw8f,vw8g}
or the HTL theory~\cite{htl_old,higgswidth}
can be used for computing the friction coefficient affecting
the Higgs motion.
But at the same time, $h$ is a short-distance
mode from the hydrodynamics point of view. Therefore, properly
accounting for the Higgs dynamics permits us to determine 
the entropy production across the bubble wall. 

A scalar field in the regime of 
\eq\nr{hierarchy2} can be described by the Langevin equation, 
\be
 h^{,\mu}_{,\mu} - V^{ }_{,h} 
 \; \simeq \;
 \frict^{ } u^\mu_{ } h^{ }_{,\mu} 
 - \varrho^{ }
 \;, \quad
  \bigl\langle\,
  \varrho^{ } (\X) \, \varrho^{ } (\Y) 
 \,\bigr\rangle
 \; \simeq \; 
 \Omega^{ } \, \delta^{(4)}_{ }(\X-\Y)
 \;. \la{langevin_hh}
\ee
The friction ($\frict^{ }_{ }$) 
and noise ($\Omega^{ }_{ }$) 
contain the effects of short-distance 
modes that have been integrated out,
with the friction transporting 
energy from $h$ to the short-distance modes, and the noise
returning it, however with minimal information content. 
Some details
on how to justify this formalism from the original field theory
are given in appendix~A. 
The Langevin description 
can resolve the bubble wall structure
on the distance scale $\lwall \sim 1/m^{ }_{\aS}$,
where hydrodynamics is not applicable. 

The coefficients $\frict^{ }_{ }$ and $\Omega^{ }_{ }$
in \eq\nr{langevin_hh} depend on $T$, 
which is constant on the microscopic length scale $\lwall$, 
as well as on $h$, which may vary. 
In addition, they could involve 
the covariant gradients $u\cdot\partial$ and $\partial\cdot\partial$. 
The coefficients $\frict^{ }_{ }$ and $\Omega^{ }_{ }$
are related to each other via fluctuation-dissipation theory, 
which states that given enough time, the system should thermalize
towards a local temperature $T$, implying that 
(cf.\ \eq\nr{lin_resp_2})
\be
 \Omega^{ }_{ } = 2 T \frict^{ }_{ }
 + 
 \rmO(u\cdot\partial,\partial\cdot\partial)
 \;. \la{lin_resp}
\ee

\section{Estimate of entropy production at bubble walls}
\la{se:algorithm}

\subsection{Basic equations}

In the dynamics described by
\eq\nr{langevin_hh}, 
the coefficient~$\frict$ induces dissipative effects, 
notably an overall 
shear viscosity~\cite[eq.~(3.42)]{scan}.\footnote{%
 It is an unfortunate nomenclature that $\eta$ is often used to 
 denote both the friction coefficient ($\Upsilon$)
 and the shear viscosity, which is inversely related to $\Upsilon$.
 } 
As long as the fluctuations
of $h$ are part of the equations, 
and $\Upsilon$ is assumed small so that
it induces a large shear viscosity, 
we need not insert any additional
shear viscosity by hand. Therefore the energy-momentum tensor
maintains the ideal form in \eq\nr{Tmunu_mixed}.
Equation~\nr{Tmunu_conserved}
can be combined with \eq\nr{langevin_hh} by requiring 
overall energy-momentum conservation, 
\ba
  0 & = & 
  T^{\mu\nu}_{(h),\mu}
  \; + \; 
  T^{\mu\nu}_{(r),\mu}  
  \;, \la{continuity} \\[2mm]
  T^{\mu\nu}_{(h),\mu}
  & \overset{\rmii{\nr{Tmunu_mixed}}}{=} & 
  h^{,\mu}_{,\mu} h^{,\nu}_{ } 
  \; \overset{\rmii{\nr{langevin_hh}}}{=} \; 
  \bigl(\,
  V^{ }_{,h} + \,\frict u^{\mu}_{ } h^{ }_{,\mu} - \varrho \,\bigr)
  \, h^{,\nu}_{ }
  \;, \la{Tmunu_h} \\[2mm]
  T^{\mu\nu}_{(r),\mu}  
  & \overset{\rmii{\nr{Tmunu_mixed}}}{=} & 
  \bigl( w u^{\mu}_{ } u^{\nu}_{ }\bigr)^{ }_{,\mu}
  + p^{,\nu}_{ }
  \;
  \underset{\rmii{\nr{Tmunu_h}}}{
  \overset{\rmii{\nr{continuity}}}{\equiv}}
  \; 
  \bigl(\, 
  -  V^{ }_{,h}  
  - \,\frict u^{\mu}_{ } h^{ }_{,\mu} + \varrho \,\bigr)
  \, h^{,\nu}_{ }
  \;. \la{Tmunu_r}
\ea

An expression for the source of the entropy current 
can be obtained by contracting \eq\nr{Tmunu_r} with $u^{ }_\nu$.
The first term on the left-hand side yields 
\be
 u^{ }_{\nu} \bigl( w u^{\mu}_{ } u^{\nu}_{ }\bigr)^{ }_{,\mu}
 \;
 \overset{u\cdot u \; = \; -1}{=} 
 \;
 - \bigl( w u^{\mu}_{ } \bigr)^{ }_{,\mu}
 \stackrel{w\; =\; T s}{=}  
 - T \bigl( s u^{\mu}_{ } \bigr)^{ }_{,\mu}
 - s u^{\mu}_{ } T^{ }_{,\mu}
 \;, \la{ds_1}
\ee
whereas the second term is rephrased as 
\be
 u^{ }_{\nu} p^{,\nu}_{ }
 = 
 u^{\mu}_{ } p^{ }_{,\mu}
 = 
 u^{\mu}_{ } 
 \, \bigl(\, 
  \overbrace{ p^{ }_{,\iT} }^{ s }
  T^{ }_{,\mu}
 + 
  \overbrace{ p^{ }_{,h} }^{ -V^{ }_{,h} }
  h^{ }_{,\mu}
 \, \bigr)\,
 \;. \la{ds_2}
\ee
Therefore 
\ba
 -u^{ }_\nu T^{\mu\nu}_{(r),\mu}
 & \underset{\rmii{\nr{ds_2}}}{\overset{\rmii{\nr{ds_1}}}{=}} & 
 T \bigl( s u^{\mu}_{ } \bigr)^{ }_{,\mu} 
 + \cancel{ V^{ }_{ ,h} } 
 u^{\mu}_{ } h^{ }_{,\mu} 
 \; 
 \overset{\rmii{\nr{Tmunu_r}}}{=}
 \; 
 \bigl(\, 
   \cancel{ V^{ }_{,h} }
  + \frict u^{\mu}_{ } h^{ }_{,\mu} - \varrho
 \,\bigr)\,
 u^{\nu}_{ } h^{ }_{,\nu} 
 \la{dsmu} \\[2mm]
 \Rightarrow \; 
 T \bigl( s u^{\mu}_{ } \bigr)^{ }_{,\mu} 
 & = & 
 \bigl(\, 
   \frict u^{\mu}_{ } h^{ }_{,\mu} - \varrho
 \,\bigr)\,
 u^{\nu}_{ } h^{ }_{,\nu} 
 \;. \la{source}
\ea
Physically, $\Upsilon$ adds to the source of the entropy 
current, because it converts 
a definite configuration of~$h$ to random thermal motion, whereas
$\varrho$ comes with a negative sign, because it describes
energy transfer away from the thermal plasma.

Subsequently, we write the scalar field as a sum of a background 
and a fluctuation, 
\be
 h = \bar{h} + \delta h 
 \;, \la{shift}
\ee
where 
$
 \langle \delta h \rangle = 0
$. 
Inserting this in \eq\nr{langevin_hh} and inspecting 
the average and the first-order fluctuation parts separately, 
with $\varrho$ assumed to be of the same magnitude as $\delta h$, 
we find the background equation
\ba
 \bar{h}^{,\mu}_{,\mu} - \bar{V}^{ }_{,h} & = &  
 \bar{\frict}^{ }_{ } u^\mu_{ } \bar{h}^{ }_{,\mu} 
 \;, \la{langevin_bg_h}  
\ea
as well as a linear differential equation for the fluctuations, 
\ba
 \bigl(\, 
    \partial^{\mu}_{ }\partial^{ }_\mu
  - \bar{V}^{ }_{,hh}
  - \bar\frict^{ }_{ } u^\mu_{ } \partial^{ }_{\mu} 
  -  \bar\frict^{ }_{,h} u^\mu_{ } \bar{h}^{ }_{,\mu}
 \,\bigr)\, \delta h 
 & = &  
 - \varrho^{ }_{ }
 \;. \la{langevin_del_h} 
\ea
Here the notation indicates that 
$\bar{V}$, $\bar\frict$ and the noise autocorrelator
$\bar\Omega$ are evaluated with $\bar{h}$ 
as argument (the flow velocity $u^{\mu}_{ }$ is also 
a background quantity). 

We recall in passing that the differential
operator acting on $\delta h$ 
in \eq\nr{langevin_del_h} 
has non-trivial zero modes, namely 
$
 \delta h \propto 
 \bar{h}^{ }_{,\nu}
$ 
if $\bar{h}$ 
depends on $\X^{\nu}_{ }$. This   
implies that the differential operator 
is not invertible in the full functional space. 
Nevertheless a special solution can be found by making use of 
a retarded Green's function, defined via
\ba
 \left(\,
 - \, 
 \partial^{\mu}_{ }\partial^{ }_\mu
 + 
 \bar V^{ }_{,hh}
 + 
 \bar\frict^{ }_{ } u^\mu_{ } \partial^{ }_{\mu} 
 + 
 \bar\frict^{ }_{,h} u^\mu_{ } \bar{h}^{ }_{,\mu} 
 \right)\,
 G^{ }_\R(\X,\Y)
 & = & 
 \delta^{(4)}_{ }(\X - \Y)
 \;, \nn[3mm] 
 G^{ }_\R \bigr|^{ }_{x^0_{ } < y^0_{ }} & = & 0
 \;, \la{G_R}
\ea
where
$
 \X \equiv (x^0_{ },\vec{x})
$.
Denoting 
$
 \int_\Y \equiv \int \! {\rm d}y^0_{ } \int \! {\rm d}^3_{ }\vec{y} 
$,
the special solution reads
\be
 \delta h  (\X) 
 \; =  \;
 \int_\Y G^{ }_\R(\X,\Y)\, 
 \varrho(\Y)
 \;. \la{h_soln}
\ee

%
\subsection{Contribution of scalar fluctuations to homogeneous entropy density}

In order to have a first impression about the effects that the
fluctuations have, we place ourselves in a homogeneous phase, 
with $ u^{\mu}_{ }\bar h^{ }_{,\mu}  = 0 $, and consider
the expectation value of 
$
 T^{\mu\nu}_\rmi{ideal}
$ from \eq\nr{Tmunu_mixed}, 
including the contribution from $\delta h$. 

Let us choose coordinates so that the $z$-axis is parallel to the direction
of the flow velocity, like in \fig\ref{fig:profile} and \eq\nr{velocity}, 
such that $u^z_{ } = - \gamma v$. Then, according to \eq\nr{Tmunu_mixed}, 
\be
 \bigl\langle\, T^{0z}_\rmi{ideal} \,\bigr\rangle 
 \; = \; 
 \underbrace{
 \bigl\langle\, \delta h^{,0}_{ } \delta h^{,z}_{ } \,\bigr\rangle 
 }_{ \,\equiv\; \langle\; T^{0z}_{ }\,\rangle^{ }_{\delta h}}
 \; + \;
 T s\, u^0_{ } u^z_{ }
 \;. \la{eff_s}
\ee

By inserting 
$
 \delta h(\X) = \int_\Y G^{ }_\R(\X,\Y) \varrho(\Y)
$
from \eq\nr{h_soln}, followed by the noise autocorrelator from 
\eqs\nr{langevin_hh} and \nr{lin_resp}, we find
\be
 \langle T^{0z}_{ }\rangle^{ }_{\delta h} 
 \; 
 \underset{\rmii{\nr{langevin_hh},\nr{lin_resp}}}{
 \overset{\rmii{\nr{eff_s},\nr{h_soln}}}{ \approx }} 
 \; 
  2 T \bar\frict \int_\Y \partial^{0}_{ } G^{ }_\R(\X,\Y)\,
  \partial^{z}_{ } G^{ }_\R(\X,\Y)
 \;. \la{T0z_1}
\ee
The Green's function from \eq\nr{G_R} can be solved with 
a Fourier transformation, 
\be
  G^{ }_\R  (\X,\Y) 
 \; \overset{\rmii{\nr{G_R}}}{=} \; 
 \int_\K
 e^{i\K\cdot(\X-\Y)}_{ } G^{ }_\R(\K)
 \;, \quad
 G^{ }_\R(\K)
 \; \equiv \;  
 \frac{1}
 {\K^2_{ } + m^2_{ } + \bar\frict\, i u\cdot\K }
 \;, \quad
 m^2 \; \equiv \; 
 \bar V^{ }_{,hh}
 \;, \la{G_R_P}
\ee
where the integration measure is defined so that 
$
 \int_\K e^{i\K\cdot(\X - \Y)}_{ } = \delta^{(4)}_{ }(\X - \Y)
$.
This yields
\be
  \langle T^{0z}_{ }\rangle^{ }_{\delta h} 
 \; 
 \underset{\rmii{\nr{G_R_P}}}{
 \overset{\rmii{\nr{T0z_1}}}{\approx}} 
 \; 
  2 T \int_\K^{ } \frac{ \bar\Upsilon k^0_{ }k^z_{ }}
                 {(\K^2_{ } + m^2_{ })^2_{ } + \bar\frict^2 (u\cdot\K)^2_{ } }
 \;. \la{T0z_2}
\ee
In order to carry out the integral, it is helpful to change variables as 
\be
  \left(\! \begin{array}{c}
  k^0_{ } \\ k^z_{ } \\ \vec{k}^{ }_\perp
 \end{array} \!\right)
 \; \equiv \; 
 \left(\!
 \begin{array}{ccc}
   \gamma & -\gamma v & 0  \\
   -\gamma v & \gamma & 0  \\ 
   0  &  0  & \mathbbm{1}^{ }_{2\times 2}
  \end{array}
 \!\right)
 \left(\!
   \begin{array}{c}
    q^0_{ } \\
    q^z_{ } \\ 
    \vec{q}^{ }_\perp 
   \end{array} \!\right)
 \;, \quad
 \left(\! \begin{array}{c}
  q^0_{ } \\ q^z_{ } \\ \vec{q}^{ }_\perp
 \end{array} \!\right)
 \; = \; 
 \left(\!
 \begin{array}{ccc}
   \gamma & \gamma v & 0  \\
   \gamma v & \gamma & 0  \\ 
   0  &  0  & \mathbbm{1}^{ }_{2\times 2}
  \end{array}
 \!\right)
 \left(\!
   \begin{array}{c}
    k^0_{ } \\
    k^z_{ } \\ 
    \vec{k}^{ }_\perp 
   \end{array} \!\right)
 \;,
 \la{lorentz}
\ee
which effectively means that we boost to the plasma rest frame.
Then \eq\nr{T0z_2} turns into 
\be
  \langle T^{0z}_{ }\rangle^{ }_{\delta h} 
 \; 
 \underset{\rmii{\nr{lorentz}}}{
 \overset{\rmii{\nr{T0z_2}}}{\approx}} 
 \; 
  2 T u^0_{ }u^z_{ }
 \int_{\Q} 
 \frac{ \bar\Upsilon \,[  (q^0_{ })^2_{ } + (q^z_{ })^2_{ } \,] }
                 {[ (q^0_{ })^2_{ } - (q^z_{ })^2_{ }
                    - \epsilon^2_{ q_\perp^{ }}]^2_{ }
                    + \bar\frict^2 (q^0_{ })^2_{ } }
 \;, \quad
  \epsilon_{q^{ }_\perp}^2 \;\equiv\; q^2_\perp + m^2
 \;. \la{T0z_3}
\ee
The integral over $q^0_{ }$ can be carried out 
with Cauchy's theorem,\footnote{%
 The results read:
 \be
  \int_{-\infty}^{\infty}
  \! \frac{{\rm d}q^0_{ }}{2\pi} \,
  \frac{1}{[(q^0_{ })^2_{ } - \epsilon^2_q]^2 
   + \bar\Upsilon^2_{ } (q^0_{ })^2_{ } } = 
  \frac{1}{2\bar\Upsilon \epsilon^2_q }
  \;, \quad 
  \int_{-\infty}^{\infty}
  \! \frac{{\rm d}q^0_{ }}{2\pi} \,
  \frac{ (q^0_{ })^2_{ } }{[(q^0_{ })^2_{ } - \epsilon^2_q]^2 
   + \bar\Upsilon^2_{ } (q^0_{ })^2_{ } } = 
  \frac{1}{2\bar\Upsilon }  
  \;. \la{cauchy}
 \ee
 } 
yielding finally
\be
  \bigl\langle\, T^{0z}_{ } \,\bigr\rangle^{ }_{\delta h} 
 \;
 \overset{\rmii{\nr{T0z_3}}}{\approx}  
 \;
 T u^0_{ }u^z_{ } 
 \underbrace{
 \int_\vec{q}^{(\Lambda)}
 \biggl(\, 1 + \frac{q_z^2 }{\epsilon_q^2} \,\biggr)
 }_{ \;\equiv\; s^{ }_{\delta h}}
 \;, \quad
 \epsilon_q^2 \;\equiv\; q^2_{ } + m^2_{ }
 \;. \la{Lambda}
\ee
Here we realized {\it a posteriori} that the spatial
momentum integral is UV divergent, and denoted by 
``$(\Lambda)$'' that it needs to be regularized. 

It is possible to crosscheck 
$s^{ }_{\delta h}$ in \eq\nr{Lambda} by comparing it with 
kinetic theory. We introduce a phase-space distribution, $f$, which
is a function of the space-time location, $\X$,
and four-momentum, $\K = (k^0_{ },\vec{k}^{ }_\perp, k^z_{ })$. 
In equilibrium, there is no dependence on~$\X$, and 
we write $f$ as 
\be
 f^{ }_0 (\K) 
 \; \equiv \;
 4 \pi\, \delta( \K^2_{ } + m^2_{ } ) \, \theta(k^0_{ })
 \,\nB^{ }(-u\cdot \K)
 \;, 
 \la{def_f0}
\ee
where $\nB^{ }$ is the Bose distribution, 
the Dirac-$\delta$ puts the momentum on-shell, 
and the Heaviside-$\theta$ restricts us to the particle branch. 
The analogue of \eq\nr{Lambda} then reads
\ba
 \bigl\langle\, T^{0z}_{ } \,\bigr\rangle^{ }_{f^{ }_0} 
 & 
 \overset{\rmii{\nr{Tmunu_mixed_1}}}{\equiv}
 & 
 \int^{ }_{\K} k^0_{ } k^z_{ } f^{ }_0(\K)
 \nn[2mm]
 & 
 \overset{\rmii{\nr{lorentz}}}{=} 
 & 
 \int^{ }_{\Q}
 u^{0}_{ } u^z_{ }
 \bigl[\, (q^0_{ })^2_{ } + (q^z_{ })^2_{ } \,\bigr]
 \,  4 \pi\, \delta( \Q^2_{ } + m^2_{ } ) \, \theta(q^0_{ }) 
 \,\nB^{ }(q^{0}_{ })
 \nn[2mm]
 &
 \overset{\rmii{\nr{def_f0}}}{=}
 &
 u^0_{ }u^z_{ }
 \underbrace{
 \int_\vec{q}
 \biggl( \epsilon^{ }_q + \frac{q_z^2}{\epsilon^{ }_q} \biggr) 
 \, \nB^{ }(\epsilon^{ }_q)
 }_{\;\equiv\; T s^{ }_{f_0} }
 \;.
 \la{T0z_5}
\ea
Comparing with \eq\nr{Lambda}, we find agreement if we define
\be
 \int_\vec{q}^{(\Lambda)} \,( \cdots \,)
 \; \equiv \; 
 \int \! \frac{{\rm d}^3_{ }\vec{q}}{(2\pi)^3_{ }} \,
 \underbrace{
 \frac{\epsilon^{ }_q \nB^{ }(\epsilon^{ }_q)}{T}
 }_{ \approx \; 1 \; \mathrm{for} \; \epsilon^{ }_q \; \ll \; T }
 \,( \cdots \,)
 \;. \la{int_q}
\ee

To summarize, the contribution to the entropy density from 
scalar field fluctuations (cf.\ \eq\nr{Lambda}) agrees with
what we expect from kinetic theory (cf.\ \eq\nr{T0z_5}). However, 
the hydrodynamic contribution is not well-defined in the UV, 
reflecting the Rayleigh-Jeans divergence of classical field theory. 
In momentum space, the issue can easily be rectified at 1-loop
level, just by defining the integration measure according
to \eq\nr{int_q}. 

\subsection{Expression for the source of the entropy current}

We now return to entropy generation, 
and insert \eq\nr{shift} into \eq\nr{source}. 
Omitting first-order terms whose expectation value vanishes, and
going up to second order, we find
\ba
 T \bigl( s u^{\mu}_{ } \bigr)^{ }_{,\mu} 
 & \simeq &
 \overbrace{ 
 \bigl[\, 
   \bar \frict +  
    \tfr12
    \bar\frict^{ }_{,hh} 
 \, \bigl\langle\, ( \delta h )^2_{ } \,\bigr\rangle\,
 \,\bigr]  
 }^{
 \; \equiv \; \frict^{ }_\rmii{eff}
 }
 \bigl(\, 
   u^{\mu}_{ } \bar h^{ }_{,\mu} 
 \,\bigr)^2_{ }
 \nn[3mm]
 & + & 
 \Bigl\langle\, 
 \underbrace{ 
 \bigl[\, 
  \bigl(\, 
   \bar\frict u^{\mu}_{ }\partial^{ }_\mu 
  + 
   \bar\frict^{ }_{,h} u^{\mu}_{ } \bar h^{ }_{,\mu}
  \,\bigr) \, \delta h 
    - \varrho
 \,\bigr]
 }_{ 
  \stackrel{\rmii{ \nr{langevin_del_h} }}{=} 
  (\, 
    \partial^{\mu}_{ }\partial^{ }_\mu - \bar{V}^{ }_{,hh}
  \,)\, \delta h
 }
 \,
 u^{\nu}_{ } \delta h^{ }_{,\nu}
 \, \Bigr\rangle
 \;. \la{source_del}
\ea
There are thus two effects at second order. 
The first is a ``renormalization'' of $\bar\frict$ 
into $\bar\frict^{ }_\rmi{eff}$, present if $\frict$ depends at least 
quadratically on $h$.
The second effect 
includes an interplay between friction and 
fluctuations. We show in momentum space
(cf.\ \se\ref{ss:homog}) that this contribution vanishes
in a homogeneous phase. However, 
in an inhomogeneous situation it gives 
a non-vanishing contribution
(cf.\ \se\ref{ss:disc}).

Inserting \eq\nr{h_soln} into \eq\nr{source_del} and averaging over
the noise according to \eq\nr{langevin_hh}, yields 
\ba
 T \bigl( s u^{\mu}_{ } \bigr)^{ }_{,\mu}(\X)
 & 
 \underset{\rmii{\nr{h_soln},\nr{langevin_hh}}}{
 \overset{\rmii{\nr{source_del}}}{\simeq}} 
 &
 \overbrace{ 
 \Bigl\{
 \bar\frict
 + 
    \frac{ 
    \bar\frict^{ }_{,hh}
    }{2} \,
 \underbrace{ 
 \bar\Omega 
 \int_\Y
 G^{2}_\R(\X,\Y)
 }_{
  \; \equiv \; \O^{ }_1(\X)
 }
 \Bigr\} 
 }^{ 
 \bar\frict^{ }_\rmii{eff}
 }
 \, 
 \bigl(\, 
   u^{\mu}_{ } \bar h^{ }_{,\mu} 
 \,\bigr)^2_{ }(\X)
 \la{eta_eff} \\[2mm]
 &   & \; + \, 
 \underbrace{ 
 \bar\Omega 
 \int_\Y
 \bigl\{\, 
  (\, 
    \partial^{\mu}_{ } \partial^{ }_\mu - \bar{V}^{ }_{,hh}
  \,) \, G^{ }_{\R}(\X,\Y)
 \bigr\}\,
 u^{\nu}_{ } \partial^{ }_\nu 
 G^{ }_\R(\X,\Y)
 }_{
 \; \equiv \; \O^{ }_2(\X) 
 }
 \;, \hspace*{5mm}  \la{source_G_R}
\ea
where the derivatives act on the first argument of $G^{ }_\R$.

\subsection{Evaluation of the source in a homogeneous phase}
\la{ss:homog}

In order to get a first impression about the structures found, 
we evaluate \eqs\nr{eta_eff} and \nr{source_G_R} in a homogeneous 
background, with 
$
  u^\mu_{ }\bar{h}^{ }_{,\hspace*{0.3mm}\mu} = 0
$. 
Inserting the Green's function from \eq\nr{G_R_P}
and $\bar\Omega$ from \eq\nr{lin_resp}, we obtain 
\ba
 \O^{ }_1(\X) 
 & 
 \underset{\rmii{\nr{G_R_P},\nr{lin_resp}}}{
 \overset{\rmii{\nr{eta_eff}}}{=}} 
 & 
 2 T \int_\K^{ } \frac{ \bar\Upsilon }
                 {(\K^2_{ } + m^2_{ })^2_{ } + \bar\frict^2 (u\cdot\K)^2_{ } }
 \;, \la{O_1} \\ 
 \O^{ }_2(\X) 
 &
 \underset{\rmii{\nr{G_R_P},\nr{lin_resp}}}{
 \overset{\rmii{\nr{source_G_R}}}{=}} 
 & 
 2 T
 \int_\K^{ } \frac{(\K^2_{ } + m^2_{ })\,\bar\Upsilon \, i u\cdot\K }
                 {(\K^2_{ } + m^2_{ })^2_{ } + \bar\frict^2 (u\cdot\K)^2_{ } }
 \;. \la{O_2}
\ea
In both cases, the dependence on the space-time 
coordinate $\X$ has dropped out. 

By power counting, \eq\nr{O_2} is UV-divergent. However, if we assume
that the integration is regularized, like in \eq\nr{Lambda}, then
the result vanishes exactly, 
due to the integrand's antisymmetry in $\K\to -\K$.

Similarly, \eq\nr{O_1} is UV-divergent by power counting. 
We may learn more about it by carrying out the integral
over $k^0_{ }$ by closing contours. Making use of
\eqs\nr{lorentz} and \nr{cauchy}, 
we find a result related to \eq\nr{Lambda},  
\be
 \O^{ }_1(\X)
 \; = \; 
 \int^{(\Lambda)}_{\vec{q}}
 \frac{T}{\epsilon_q^2}
 \; \overset{\rmii{\nr{int_q}}}{=} \; 
 \int \! \frac{{\rm d}^3_{ }\vec{q}}{(2\pi)^3_{ }} \,
 \frac{\nB^{ }(\epsilon^{ }_q)}{\epsilon^{ }_q }
 \;. \la{O_1_b}
\ee
The integral is nothing but the  basic thermal ``tadpole'' integral, 
\be
 \int^{ }_{\vec{k}} \frac{\nB^{ }(\epsilon^{ }_k)}{\epsilon^{ }_k}
 \; \overset{m\;\ll\;T}{\approx} \; 
 \frac{T^2_{ }}{12}
 \;, \la{tadpole}
\ee
representing a finite ``renormalization'' of 
$\bar\Upsilon$ through scalar field fluctuations. 

\subsection{Discontinuity of entropy current across a bubble wall}
\la{ss:disc}

In the presence of a bubble wall, 
the term in \eq\nr{eta_eff}, 
containing the background variables, 
gives a positive contribution to the source of the entropy current
(if $\bar\Upsilon^{ }_\rmi{eff} > 0$). 
Here we focus on the effects from scalar field fluctuations,  
contained in $\O^{ }_2(\X)$
from \eq\nr{source_G_R}. We assume a geometry like in 
\fig\ref{fig:profile}, in which the $z$-direction takes a special role. 
The situation is considered stationary (i.e.\ with a time-independent
flow in the negative $z$-direction), 
so that we may Fourier transform in 
the temporal and transverse spatial directions. Then 
\be
 G^{ }_\R(\X,\Y)
 \;
 \overset{\rmii{\nr{G_R_P}}}{=} 
 \;
 \int_{(\omega,\vec{k}^{ }_\perp)} 
 \, e^{-i \omega (x^0_{ } - y^0_{ }) + 
        i \vec{k}^{ }_\perp\cdot(\vec{x}^{ }_\perp - \vec{y}^{ }_\perp)}_{ }
 \, 
 G^{ }_\R(z,y^3_{ },\cdot)
 \;, \quad \omega \; \equiv \; k^0_{ }
 \;, \quad z \; \equiv \; x^3_{ }
 \;, \la{mixed_rep}
\ee
where 
$\cdot \equiv \omega,\vec{k}^{ }_\perp$ stands
for variables that are not shown explicitly. Furthermore, 
we denote $\X \to z$. 
Then \eq\nr{G_R} gets replaced with 
\be
 \Bigl[\, 
 - \partial_z^2 
 + \bar\frict\,\bigl(  u^z_{ } \partial^{ }_z 
 - i u^0_{ }\omega \,\bigr) 
 \; 
 \underbrace{ 
 - \omega^2 + 
 \hspace*{-3mm}
 \overbrace{ k_\perp^2 + \bar{V}^{ }_{,hh}
  }^{
 \equiv\; k^2_\perp + m^2_{ }(z) \; \equiv \; \epsilon^2_{k_\perp}}
 \hspace*{-3mm}
 }_{\;\equiv\; \Delta m^2_{ }(z) }
 + \bar\Upsilon^{ }_{,h} u^z_{ } \bar{h}^{ }_{,z}
   \,\Bigr]\,
 G^{ }_\R(z,y^3_{ },\cdot) 
 \;
 \underset{\rmii{\nr{mixed_rep}}}{
 \overset{\rmii{\nr{G_R}}}{=}}
 \; 
 \delta(z - y^3_{ })
 \;. \la{G_R_z}
\ee
We remark that in the spatial directions, the two
boundary conditions required for solving the second-order
equation amount to 
\be
 \lim_{z_{ }\to \pm\infty} G^{ }_\R(z,y^3_{ },\cdot)
 \; = \; 0
 \;, \la{bc}
\ee 
because by causality no information can travel infinitely
far in a finite period of time. 

With the help of the partially Fourier-transformed 
Green's function from \eq\nr{mixed_rep}, 
and inserting 
$\bar\Omega  \approx 2 T \bar\Upsilon$ from \eq\nr{lin_resp}, 
the observable from \eq\nr{source_G_R} becomes
\be
 \O^{ }_2(z)
 \;
 \underset{\rmii{\nr{lin_resp}}}{
 \overset{\rmii{\nr{source_G_R}}}{\approx}}
 \;
 2 T \bar\frict
 \int_{(\omega,\vec{k}^{ }_\perp)} 
 \int_{-\infty}^{\infty} \! {\rm d}y^3_{ } \, 
 (\partial_z^2 - \Delta m^2)
  G^{ }_\R (z,y^3_{ },\cdot)
 (i u^0_{ }\omega + u^z_{ } \partial^{ }_z)
  G^{*}_\R (z,y^3_{ },\cdot)
 \;. \la{O_2_z} 
\ee

In order to make progress, we realize that 
the replacement $\omega\to -\omega$ corresponds to complex
conjugation of $G^{ }_\R$ (cf.\ \eq\nr{G_R_z}). We write
\be
 \int_\omega \, f(\omega) 
 \; = \; 
 \int_{\omega^+_{ }} 
 \, \bigl[\, f(\omega) + f(-\omega) \,\bigr]
 \;, \quad
 \int_{\omega^+_{ }} 
 \; \equiv \; 
 \int_0^\infty \! {\rm d}\omega 
 \;. \la{symmetry}
\ee
Omitting arguments and subscripts from $G^{ }_\R$; 
and recalling from \fig\ref{fig:profile} that the 
hydrodynamic variables $T$ and $v$ can be treated
as constants on the distance scales at which the 
wall profile varies, we then obtain 
(recalling $u^z_{ } = - u^0_{ } v$)
\ba
 \O^{ }_2(z)
 &
 \underset{\rmii{\nr{symmetry}}}{
 \overset{\rmii{\nr{O_2_z}}}{\approx}} 
 &
 2 T \bar\frict u^0_{ }
 \int_{(\omega^+_{ },\vec{k}^{ }_\perp)} 
 \int_{-\infty}^{\infty} \! {\rm d}y^3_{ } \, 
 \Bigl\{\, 
 \overbrace{
 (\partial_z^2 G^{ }_\R)\, i\omega\, G^{*}_\R 
 -  
 (\partial_z^2 G^{*}_\R)\, i\omega\, G^{ }_\R 
 }^{  
 i \omega \partial^{ }_z 
 (
 G^{*}_R \partial^{ }_z G^{ }_\R 
 - 
 G^{ }_R \partial^{ }_z G^{*}_\R 
 )
 } \qquad \mbox{(i)}
 \la{O_2_a}  \\[2mm]
 & & 
 \hspace*{4cm}
 \overbrace{
 - 
 (\partial_z^2 G^{ }_\R)\, v\, \partial^{ }_z G^{*}_\R 
 -  
 (\partial_z^2 G^{*}_\R)\, v\, \partial^{ }_z G^{ }_\R 
 }^{  
 - v\, \partial^{ }_z 
 ( \partial^{ }_z G^{ }_\R \partial^{ }_z G^{*}_\R 
 )
 } \qquad \mbox{(ii)}
 \nn[2mm]
 & & 
 \hspace*{4cm}
 - \cancel{ \Delta m^2_{ } G^{ }_\R \, i\omega \, G^{*}_\R }
 + \cancel{ \Delta m^2_{ } G^{*}_\R \, i\omega \, G^{ }_\R }
 \nn[2mm]
 & & 
 \hspace*{4cm}
 \underbrace{
 + 
  \Delta m^2_{ } G^{ }_\R \, v\, \partial^{ }_z G^{*}_\R 
 +  
  \Delta m^2_{ } G^{*}_\R \, v\, \partial^{ }_z G^{ }_\R 
 }_{  
   \Delta m^2_{ } \, v \, 
 \partial^{ }_z 
 ( G^{ }_\R  G^{*}_\R 
 )
 }
 \, \Bigr\} 
 \;. \qquad \mbox{(iii)} \nonumber
\ea
Subsequently, we integrate the source of the entropy current
according to \eqs\nr{Delta_s_def} and \nr{source_G_R}, 
\be
 \Delta^{ }_{s u^z_{ }}
  \;
  \underset{\rmii{\nr{source_G_R}}}{
  \overset{\rmii{\nr{Delta_s_def}}}{\supset}}
  \;
 \frac{1}{T} 
 \int_{z^{ }_-}^{z^{ }_+} \! {\rm d}z \, \O^{ }_2(z)
 \;.
 \la{Delta_s_2}
\ee

Now, one might think that 
the terms in \eq\nr{O_2_a} which are total derivatives, 
vanish upon recalling the boundary conditions from \eq\nr{bc}. 
This is, however, not trivially the case, since \eq\nr{O_2_a} 
involves an integration over $y^3_{ }$, so that short relative distances
do appear. We need to determine the boundary terms explicitly. 
It helps to note from \eqs\nr{G_R_P} and \nr{mixed_rep} that, 
when we are far away from the bubble wall, 
and thus in a homogeneous phase, then 
\be
 G^{ }_\R(z,y^3_{ },\cdot)
 \;
 \underset{\rmii{\nr{G_R_P},\nr{mixed_rep}}}{
 \overset{|z|\;\gg\; m_h^{-1}}{\approx}}
 \;
 \int_{k^z_{ }} e^{i k^z_{ }( z - y^3_{ }) }_{ }
 \, G^{ }_\R (\K)
 \;. \la{inv_F}
\ee
We also denote
\be
 [ \cdots ]^+_- 
 \; \equiv \; 
 [ \cdots ](z^{ }_+) - [ \cdots ](z^{ }_-)
 \;. \la{substitutions}
\ee

Let us work out the individual contributions. 
For the first term, we obtain
\ba
 \Delta^\rmi{(i)}_{s u^z_{ }}
 & 
 \underset{\rmii{\nr{Delta_s_2}}}{
 \overset{\rmii{\nr{O_2_a}}}{=}} 
 & 
 2 \bar\Upsilon u^0_{ }
 \int_{(\omega^+_{ },\vec{k}^{ }_\perp)} 
 \int_{-\infty}^{\infty} \! {\rm d}y^3_{ } \,
 i \omega  
 \bigl[\;
  G^{*}_\R \partial^{ }_z G^{ }_\R
  - 
  G^{ }_\R \partial^{ }_z G^{*}_\R
  \;\bigr]^+_-
 \nn[2mm] 
 &
 \underset{\rmii{\nr{inv_F}}}{
 \overset{\rmii{\nr{symmetry}}}{\approx}}
 & 
 2 \bar\Upsilon u^0_{ }
 \int_\K (- k^0_{ } k^z_{ })
 \bigl[\; G^{ }_\R(\K) G^{*}_\R(\K) \;\bigr]^+_-
 \nn[2mm]
 & 
 \overset{\rmii{\nr{lorentz}}}{=}
 & 
 2 \bar\Upsilon u^z_{ }
 \int_\Q 
 \biggl\{ 
 \frac{(-\gamma^2_{ })[ (q^0_{ })^2_{ } + (q^z_{ })^2_{ }]}
              {[(q^0_{ })^2 - \epsilon_q^2]^2_{ }
                + \bar\Upsilon^2_{ } (q^0_{ })^2 }
 \biggr\}^+_-
 \nn[2mm]
 & 
 \overset{\rmii{\nr{cauchy}}}{=} 
 &  
 u^z_{ }
 \int^{(\Lambda)}_\vec{q}
 \biggl\{ 
 (-\gamma^2_{ }) \biggl[ 1 
  + \frac{(q^z_{ })^2_{ }}{\epsilon_q^2} \biggr]
 \biggr\}^+_-
 \;. \la{Delta_s_i}
\ea
For the second term, we find
\ba
 \Delta^\rmi{(ii)}_{s u^z_{ }}
 & 
 \overset{-u^0_{ }v = u^z_{ }}{=} 
 & 
 2 \bar\Upsilon u^z_{ }
 \int_{(\omega^+_{ },\vec{k}^{ }_\perp)} 
 \int_{-\infty}^{\infty} \! {\rm d}y^3_{ } \, 
 \bigl[\; \partial^{ }_z G^{ }_\R \partial^{ }_z G^{*}_\R \;\bigr]^+_-
 \nn[2mm] 
 &
 \underset{\rmii{\nr{inv_F}}}{
 \overset{\rmii{\nr{symmetry}}}{\approx}}
 & 
 \bar\Upsilon u^z_{ }
 \int_\K (k^z_{ })^2_{ }
 \bigl[\; G^{ }_\R(\K) G^{*}_\R(\K) \;\bigr]^+_-
 \nn[2mm]
 & 
 \overset{\rmii{\nr{lorentz}}}{=}
 & 
 \bar\Upsilon u^z_{ }
 \int_\Q 
 \biggl\{ 
 \frac{\gamma^2_{ }[v^2_{ } (q^0_{ })^2_{ } + (q^z_{ })^2_{ }]}
              {[(q^0_{ })^2 - \epsilon_q^2]^2_{ }
                + \bar\Upsilon^2_{ } (q^0_{ })^2 }
 \biggr\}^+_-
 \nn[2mm]
 & 
 \overset{\rmii{\nr{cauchy}}}{=} 
 &  
 \frac{u^z_{ }}{2}
 \int^{(\Lambda)}_\vec{q}
 \biggl\{ 
 \gamma^2_{ }\biggl[ v^2_{ }
  + \frac{(q^z_{ })^2_{ }}{\epsilon_q^2} \biggr]
 \biggr\}^+_-
 \;. \la{Delta_s_ii}
\ea
The last term yields
\ba
 \Delta^\rmi{(iii)}_{s u^z_{ }}
 & 
 \underset{\rmii{\nr{Delta_s_2}}}{
 \overset{\rmii{\nr{O_2_a}}}{=}} 
 & 
 2 \bar\Upsilon u^0_{ }
 \int_{(\omega^+_{ },\vec{k}^{ }_\perp)} 
 \int_{-\infty}^{\infty} \! {\rm d}y^3_{ } \,
 \int_{z^{ }_-}^{z^{ }_+} \! {\rm d}z_{ } \,
 \Delta m^2_{ } \, v \, 
 \partial^{ }_z 
 \bigl[\;
  G^{*}_\R  G^{ }_\R
  \;\bigr]
 \nn[2mm] 
 & 
 \underset{\rmii{\nr{symmetry}}}{
 \overset{\rmii{\nr{G_R_z}}}{=}} 
 & 
 \overbrace{
 \bar\Upsilon u^z_{ }
 \int_{(\omega,\vec{k}^{ }_\perp)} 
 \int_{-\infty}^{\infty} \! {\rm d}y^3_{ } \,
 \Bigl\{\; [(k^0_{ })^2 - \epsilon^2_{k_\perp}]
 \; G^{*}_\R  G^{ }_\R \;\Bigr\}^+_- 
 }^{ \equiv \Delta_{s u^z_{ }}^\rmii{(iii/a)}}
 \nn[2mm]
 & & 
 \; + \; 
 \underbrace{
 \bar\Upsilon u^z_{ }
 \int_{z^{ }_-}^{z^{ }_+} \! {\rm d}z_{ } \,
 \bigl(\, 
 \partial^{ }_z 
 \Delta m^2_{ }  
 \,\bigr)
 \int_{(\omega,\vec{k}^{ }_\perp)} 
 \int_{-\infty}^{\infty} \! {\rm d}y^3_{ } \,
 \bigl[\;
  G^{*}_\R  G^{ }_\R
  \;\bigr]
  }_{ \equiv \Delta_{s u^z_{ }}^\rmii{(iii/b)}}
  \;, \la{Delta_s_iii_pre}
\ea
where
\ba
 \Delta_{s u^z_{ }}^\rmi{(iii/a)}
 & 
 \underset{\rmii{\nr{inv_F}}}{
 \overset{\rmii{\nr{lorentz}}}{\approx}}
 & 
 \bar\Upsilon u^z_{ }
 \int_\Q 
 \biggl\{ 
 \frac{\gamma^2_{ }[ (q^0_{ })^2_{ } + v^2_{ }(q^z_{ })^2_{ }]
      - \epsilon^2_{q_\perp}}
              {[(q^0_{ })^2 - \epsilon_q^2]^2_{ }
                + \bar\Upsilon^2_{ } (q^0_{ })^2 }
 \biggr\}^+_-
 \nn[2mm]
 & 
 \overset{\rmii{\nr{cauchy}}}{=} 
 &  
 \frac{u^z_{ }}{2}
 \int^{(\Lambda)}_\vec{q}
 \biggl\{ 
 \gamma^2_{ } \biggl[ 1 
  + \frac{v^2_{ }(q^z_{ })^2_{ }}{\epsilon_q^2} \biggr]
  - \frac{ \epsilon^2_{q_\perp} }{\epsilon_q^2}
 \biggr\}^+_-
 \;. \la{Delta_s_iii_a}
\ea
Summing together \eqs\nr{Delta_s_i}, \nr{Delta_s_ii}, 
and \nr{Delta_s_iii_a}, we get
\ba
 \Delta^{ }_{s u^z_{ }} 
 &
 \underset{\rmii{\nr{Delta_s_iii_a}}}{
 \overset{\rmii{\nr{Delta_s_i},\nr{Delta_s_ii}}}{\supset}}
 & 
 \frac{u^z_{ }}{2}
 \int^{(\Lambda)}_\vec{q}
 \biggl\{ 
 \gamma^2_{ } \biggl[1+ v^2 -2 
  + \bigl( 1 + v^2_{ }  - 2 \bigr)\frac{(q^z_{ })^2_{ }}{\epsilon_q^2}  
  \biggr]
  - \frac{ \epsilon^2_{q_\perp} }{\epsilon_q^2}
 \biggr\}^+_-
 \nn[2mm] 
 & = & 
 \frac{u^z_{ }}{2}
 \int^{(\Lambda)}_\vec{q}
 \biggl\{ 
  - 1 
  - \frac{(q^z_{ })^2_{ }}{\epsilon_q^2}  
  - \frac{ \epsilon^2_{q_\perp} }{\epsilon_q^2}
 \biggr\}^+_-
 \; 
 = 
 \; 
 u^z_{ }
 \int^{(\Lambda)}_\vec{q}
 \bigl\{ -1 \bigr\}^+_-
 \; 
 = 
 \;
 0
 \;. \la{Delta_s_disc}
\ea
Therefore, the boundary terms give no contribution. 

The contribution that did not vanish is the second term
of \eq\nr{Delta_s_iii_pre}. Undoing the Fourier representation, 
it can be manipulated as 
\ba
 \Delta_{s u^z_{ }}^\rmii{(iii/b)}
 & = & 
 \bar\frict u^z_{ }
 \int_{z^{ }_-}^{z^{ }_+} \! {\rm d}z_{ } \,
 (\partial^{ }_z  m^2_{ } ) \, 
 \int_\Y G^{ }_\R(z,\Y) G^{ }_\R(z,\Y)
 \nn[2mm] 
 & 
 \overset{\rmii{\nr{eta_eff}}}{=} 
 & 
 \frac{ u^z_{ } }{2T} 
 \int_{z^{ }_-}^{z^{ }_+} \! {\rm d}z_{ } \,
 (\partial^{ }_z m^2_{ } ) \, 
 \O^{ }_1(z)
 \nn[2mm] 
 & 
 \underset{ m \; \ll \; T}{
 \overset{\rmii{\nr{O_1_b}}}{\approx}} 
 & 
 \frac{ u^z_{ } }{2}
 \bigl[\, (m^2_{ })^+_{ } - (m^2_{ })^-_{ } \,\bigr] 
 \int_\vec{q}^{(\Lambda)} \frac{1}{\epsilon^2_{q}}
 \la{pre_Delta_s_res} \\[2mm] 
 & 
 \underset{\rmii{\nr{int_q}}}{
 \overset{u^z_{ }\, = \, -\gamma v}{\approx}} 
 & 
 \gamma v \, 
 \frac{ 
 \bigl[\, (m^2_{ })^-_{ } - (m^2_{ })^+_{ } \,\bigr] 
 }{T}
 \int_\vec{q} \frac{\nB(\epsilon^{ }_q)}{2 \epsilon^{ }_q}
 \;. \la{Delta_s_res}
\ea
Normally, masses are assumed larger in the Higgs phase 
(for the scalar particles, positive rather than possibly tachyonic), so 
the result is positive. 

The entropy contribution of \eq\nr{Delta_s_res}, 
which enters into \eq\nr{T_disc}, represents our main result. 
In the next section, we show how it can be understood in the
language of kinetic theory, and how it compares with the 
corresponding literature. 

\subsection{Comparison with kinetic theory}
\la{ss:kinetic}

In order to gain further understanding on \eq\nr{Delta_s_res}, we show
how the same result can be obtained from kinetic theory. Specifically, 
we consider a situation in which collisions play no role, reflecting
the limit $\Upsilon\to 0$ on the side of the Langevin description, and
demonstrate that \eq\nr{Delta_s_res} is nevertheless reproduced. 

In the collisionless limit, kinetic theory takes the form of the 
Liouville equation, 
\be
 \frac{{\rm d}f}{{\rm d}t}
 \; 
 = 
 \; 
 \bigl(\,
 \partial^{ }_t 
 + \vec{v} \cdot \nabla^{ }_\vec{r} 
 + \dot{\,\vec{k}} \cdot \nabla^{ }_\vec{k}
 \,\bigr) \, f
 \;
 \overset{\Upsilon\,\to\,0}{=} 
 \;
 0 
 \;. \la{liouville_1}
\ee
Following the arguments reviewed in ref.~\cite{vw1}, if we consider
a stationary situation, we can assume that particle energies are 
conserved, and thus write
\be
 \epsilon_k^2
 \; = \;
 (k^z_{ })^2_{ } + k_\perp^2 + m^2 
 \; = \; 
 \mbox{const.} 
 \;
 \overset{{\rm d}/{\rm d}t}{\Rightarrow}
 \; 
 2 k^z_{ } \dot{k}^z_{ }
 \; = \; 
 - (\partial^{ }_z m^2_{ }) \, \dot{z}
 \;. \la{energy_1}
\ee
Writing the velocity as 
$
 \dot{z} = v^z_{ } = k^{z}/\epsilon^{ }_k
$, 
leads to 
$
 \epsilon^{ }_k\, \dot{k}^z_{ } 
 = 
 - (\partial^{ }_z m^2_{ })/2
$.
Multiplying \eq\nr{liouville_1} with $\epsilon^{ }_k$ hence yields
\be
 \biggl(\, 
 k^\mu_{ }\partial^{ }_\mu - 
 \frac{\partial_z m^2_{ }}{2}
 \partial^{ }_{k^z_{ }}
 \,\biggr)
 \, f 
 \;
 \underset{\rmii{\nr{energy_1}}}{
 \overset{\rmii{\nr{liouville_1}}}{=}}
 \; 
 0
 \;. \la{liouville_2}
\ee

Let us turn to the energy-momentum tensor. In kinetic theory, 
we can express the contribution of $f$ to it as 
\be
 T^{\mu\nu}_\rmi{ideal}
 \; \supset \; 
 T^{\mu\nu}_{(f)} 
 \;, \quad
 T^{\mu\nu}_{(f)}
 \; \equiv \; 
 \int_\K k^\mu_{ }k^\nu_{ } \, f
 \;. \la{Tmunu_mixed_1}
\ee
Because of the interaction with the bubble wall, 
energy-momentum is not conserved, since
\be
 T^{\mu\nu}_{(f),\mu}
 \; 
 \overset{\rmii{\nr{Tmunu_mixed_1}}}{=} 
 \; 
 \int_\K k^{\nu}_{ }k^\mu_{ }\partial^{ }_\mu f
 \; 
 \overset{\rmii{\nr{liouville_2}}}{=} 
 \; 
 \frac{\partial^{ }_z m^2_{ }}{2}
 \int_\K k^\nu_{ } \partial^{ }_{k^z_{ }} f
 \; 
 \neq
 \; 
 0
 \;. \la{Tmunu_mixed_2}
\ee
According to \eqs\nr{ds_1} and \nr{ds_2}, the source of
the entropy current can be obtained by contracting this 
with $u^{ }_\nu$, leading to 
\be
 T (s u^\mu_{ })^{ }_{,\mu} 
 \; 
 \underset{\rmii{\nr{ds_2}}}{
 \overset{\rmii{\nr{ds_1}}}{\supset}} 
 \; 
 - u^{ }_\nu T^{\mu\nu}_{(f),\mu}
 \; 
 \overset{\rmii{\nr{Tmunu_mixed_2}}}{=} 
 \; 
 - 
 \frac{\partial^{ }_z m^2_{ }}{2}
 \int_\K u^{ }_\nu k^\nu_{ } \partial^{ }_{k^z_{ }} f
 \; 
 \underset{\rmii{integration}}{
 \overset{\rmii{partial}}{=}} 
 \; 
 \frac{u^{ }_z}{2}
 \partial^{ }_z m^2_{ }
 \int_\K f 
 \;. \la{Tmunu_mixed_3} 
\ee
By approximating $f$ with $f^{ }_0$ from 
\eq\nr{def_f0}, computing 
$\Delta^{ }_{s u^z_{ }}$ from \eq\nr{Delta_s_def}, 
and going over to the coordinate system from \eq\nr{lorentz}, 
we obtain 
\be
 \Delta^{ }_{s u^z_{ }} 
 \; 
 \underset{\rmii{\nr{Tmunu_mixed_3}}}{
 \overset{\rmii{\nr{Delta_s_def}}}{\supset}}
 \; 
 \gamma v \, \frac{(m^2_{ })^-_{ } - (m^2_{ })^+_{ }}{T}
 \int_\vec{q} \frac{\nB^{ }(\epsilon^{ }_q)}{2 \epsilon^{ }_q}
 \;, \la{Tmunu_mixed_4}
\ee
which indeed agrees with \eq\nr{Delta_s_res}.

We remark that, apart from the overall factor $\gamma v$ that 
is associated with our definition of the entropy current
(cf.\ \eq\nr{Delta_s_def}), the result in  
\eq\nr{Delta_s_res}, multiplied by $T$, 
agrees with the force per area exerted
on the bubble wall by the ultrarelativistic particles that 
pass through it, via ``$1\to 1$'' processes~\cite[eq.~(3.7)]{vw1}. 
The quantity $Ts = e + p$ involves the pressure, so this agreement
makes sense. That said, in ref.~\cite{vw1}, the corresponding
expression comprises the {\em sum over all plasma particles}. In our
approach, the influence of the other particles is supposed to be
captured by  the functions $e$ and $p$, 
as well as by the value of $\Upsilon$.

\section{Conclusions and outlook}
\la{se:concl}

Hydrodynamic simulations extended by a scalar field  
represent the state-of-the-art tool for gravitational wave 
predictions from an electroweak phase transition~\cite{simu} 
(for overall features, computationally less expensive 
methods have also been developed, in which the 
structure of the bubble wall does not need to be resolved). However, by
the fluctuation-dissipation theorem, the presence of friction
in the equation of motion of the scalar field 
necessitates the inclusion of 
fluctuations. The purpose of the current study has been to 
explore the qualitative novelties that the fluctuations can bring
with them. 

Our framework is based on a scale hierarchy
and an associated set of effective theories
(cf.\ \eq\nr{hierarchy2} and \fig\ref{fig:profile}).
As a main result, we have shown that even 
if the scalar field friction coefficient were tuned
to zero, the fluctuations generate a finite entropy discontinuity
across the bubble wall. This has a bearing
on the idea of constraining the bubble wall velocity by assuming
the absence of any entropy production
(referred to as ``local thermal equilibrium'').
Our finding implies that such an assumption gives
an {\em unsaturated} upper bound on $v^{ }_\rmi{w}$. 

Our result for the entropy discontinuity, 
\eq\nr{Delta_s_res}, 
matches the result 
of ref.~\cite{vw1} for the force per area 
caused by $1\to 1$ processes in a 
``ballistic regime'' of particle scatterings across the wall
(for a recent discussion see, e.g.,\ ref.~\cite{vw38}).
However, our expression serves a different purpose, 
namely determining the right-hand side of \eq\nr{T_disc}, 
and thereby completing the information needed for 
obtaining the hydrodynamic fluid profile.  

The underlying goal of 
our investigation is to suggest that it should be interesting
and relevant to include scalar field fluctuations in 
hydrodynamic simulations of bubble dynamics. Even if 
they add small-scale structure to the problem, 
it is computationally less expensive
to include a single fluctuating field, rather than a momentum-dependent
phase space distribution of some particle species
at each space-time location. That said, the issue
of Rayleigh-Jeans UV divergences (cf.\ \eq\nr{Lambda}) needs 
to be faced. We remark that here the production of gravitational waves
from scalar fluctuations can be turned into a useful probe, 
as the corresponding signal can be derived analytically, 
and is indeed very sensitive to the UV regularization~\cite{hydro}. 

%

%
\appendix
\renewcommand{\thesection}{\Alph{section}} 
\renewcommand{\thesubsection}{\Alph{section}.\arabic{subsection}}
\renewcommand{\theequation}{\Alph{section}.\arabic{equation}}

%
\section{Sketch for the origin of the noise autocorrelator}
\la{se:force}

Employing Euclidean coordinates ($X = (\tau,\vec{x}) \in \mathbbm{R}^4_{ }$)
in order to suppress minus signs
(after Wick rotation back 
to Minkowskian signature, $\partial^{ }_\tau \to -i \partial^{ }_t $
and 
$L^{ }_\E \to - \mathcal{L}$), 
we consider a Lagrangian whose dependence on a neutral Higgs field $h$
takes the form 
\be
 L^{ }_\E = \fr12 \partial^{ }_\mu h    \,\partial^{ }_\mu h 
          + V(h)
          + h \underbrace{J^{ }_1}_{\rmO(g)}
          + \; h^2_{ } \underbrace{J^{ }_2}_{\rmO(g^2)}
          + \; ... \;. 
\ee
Here $J^{ }_{1,2}$ are operators depending on other fields, such 
as fermions, gauge fields, or charged components of the Higgs doublet.
We now shift the scalar $h$ around the respective
background solution,  
\be
 h \to \bar h + \delta h 
 \;, \la{background}
\ee
with the background assumed to be semiclassical, i.e.\ 
\be
 \delta^{ }_h S^{ }_\E 
 \;
 \underset{  }{  
  \overset{ \rmii{$ h = \bar h $ } }{ = } } 
 \;
 0 
 \;, \quad
 S^{ }_\E
 \;
  =
 \;
 \int_X L^{ }_\E
 \;. \la{semi-clas}
\ee

The fluctuations $\delta h$
are assumed to feel two types of physics. 
The IR dynamics of $\delta h$, 
associated with the scale $m^{ }_h$ in 
\eq\nr{hierarchy2}, has been 
treated in the body of the text. The IR dynamics is 
a response to UV fluctuations, originating
from $J^{ }_{1,2}$ and carrying harder momenta
$\sim gT$, which has to be accounted for by the noise. 
In this logic, \eq\nr{semi-clas} remains true if
we include the IR fluctuations as the background fields. 
The semiclassical equations can then be written as   
\ba
 0 & = & 
 -\partial_\mu^2 ( \bar h + \delta h ) + 
 \partial^{ }_h V(\bar h + \delta h ) + J^{ }_1 + 
 2 (\bar h + \delta h ) J^{ }_2 + ... 
 \;. 
\ea
We now expand the equations to first order 
in $\delta h \ll \bar h$, to obtain
\be
     \partial_\mu^2 \bar h - \partial^{ }_h \bar V
 \; = \; 
 \left( 
     -\partial_\mu^2 + \partial_h^2 \bar V
 \right)
     \delta h
 + 
     J^{ }_1 + 2 (\bar h + \delta h) J^{ }_2 + ...
 \;. \la{eom}
\ee
As discussed around \eqs\nr{langevin_bg_h}--\nr{langevin_del_h}, 
this can be split into separate equations for 
$\bar h$ and $\delta h$, 
after we have defined the fluctuations
of $\delta h$ to originate from the noise.

We then envisage ``integrating out'' 
the fields appearing in $J^{ }_{1,2}$.
Thereby we obtain an effective equation of motion for $\delta h$. 
The leading contribution originates from $J^{ }_1$, whereas 
$J^{ }_{2}$ is of $\rmO(g^2)$.\footnote{%
 A single insertion of $J^{ }_{2}$ is of the same order as 
 two insertions of $J^{ }_1$, however a single insertion 
 leads to a momentum-independent contribution, which has 
 no imaginary part in the sense of \eq\nr{relation}.
 } 
Therefore, at leading order in the 
weak-coupling limit, it is sufficient to compute the 
contribution originating from $J^{ }_1$ to the equations of motion. 

Imposing finally detailed balanced
(i.e.,\ the energy inserted by the kicks described by 
$J^{ }_1 \simeq - \varrho$
should be compensated for by dissipation, represented by the 
friction $\frict$), 
we expect that \eq\nr{eom} turns into\footnote{%
 The overall sign of the noise term 
 $\varrho \simeq - J^{ }_1$ is not meaningful, 
 given that $\langle \varrho \rangle = 0$. However, to conform with
 the usual conventions for stochastic differential equations, we have chosen
 it to coincide with the sign of the second time derivative, so that  
 $\partial_t^2 ( \bar{ h } + \delta h) + ... 
 = ... + \varrho$.
 } 
\be
 \partial^\mu_{ }\partial^{ }_\mu 
 ( \bar{h} + \delta h ) 
 -  V^{ }_{,h} ( \bar{h} + \delta h )
 \; \simeq \;
 \frict u^\mu_{ }\partial^{ }_\mu  
 ( \bar{h} + \delta h )
 - \varrho
 \;, \la{langevin}
\ee
where we have Wick-rotated to $\X \equiv (t,\vec{x})$ for 
a clear physical picture. 
If $ \bar{h} + \delta h $ 
varies on intermediate scales
($m^{ }_h \ll gT$), 
the correlations of the force $J^{ }_1$ are on 
short length and time scales in comparison. 
Then we may assume the noise to be white, 
\be
 \bigl\langle\,
  \varrho^{ } (\X) \, \varrho^{ } (\Y) 
 \,\bigr\rangle
 \; \simeq \; 
 \Omega^{ } (\X) \, \delta^{(4)}_{ }(\X-\Y)
 \;. \la{noise}
\ee

As a final step, 
we note from \eq\nr{noise} that 
the autocorrelator of the noise 
can be obtained from a 2-point function. 
This can subsequently be matched onto a corresponding
2-point function in the quantum theory containing $J^{ }_1$, 
\be
 \Omega^{ } (\X) = 
 \int_{\Y} 
 \bigl\langle\,
  \varrho^{ } (\X) \, \varrho^{ } (\Y) 
 \,\bigr\rangle
 \; \simeq \; 
 \lim_{\omega\to 0} 
 \underbrace{
 \int_{\Y} e^{i\omega(x^0_{ } - y^0_{ })}
 \Bigl\langle\,
  \frac{1}{2} \bigl\{ 
  J^{ }_1(\X) , J^{ }_1(\Y) 
 \bigr\}
 }_{ \equiv \Delta(\omega;\X)}
 \,\Bigr\rangle
 + \rmO(g^4)
 \;. \la{Omega_def}
\ee
Here a frequency has been introduced as a calculational
tool (computations are often easier with $\omega > 0$). Moreover we have 
adopted the symmetric operator ordering, given that it has a classical limit.  
In local thermal equilibrium, the symmetric operator ordering is related to 
other orderings, in particular
\be
 \Delta(\omega;\X) = \bigl[ 1 + 2 \nB^{ }(\omega) \bigr]\, 
 \im G^{ }_\R(\omega;\X)
 \;, \la{relation}
\ee
where 
$
 \nB^{ }(\omega) \equiv 1/( e^{\omega/T}_{ } - 1 )
$ is the Bose distribution, 
and $G^{ }_\R$ is a retarded Green's function. 
Defining
\be
 \frict^{ } (\X) \; \equiv \; \lim^{ }_{\omega \to 0} 
 \frac{\im G^{ }_\R (\omega;\X)}{\omega} 
 \;,
\ee
and going to the IR limit $\omega \ll \pi T$
(cf.\ \eqs\nr{hierarchy2} and \nr{lin_resp}), 
\eqs\nr{Omega_def} and \nr{relation} then yield 
the fluctuation-dissipation relation,
\be
 \Omega^{ }
 \; 
 \approx
 \;
  2 T \frict^{ } 
 \;. \la{lin_resp_2}
\ee

\small{
%

}


\begin{thebibliography}{99}

\bibitem{lisa_pt}
  C.~Caprini {\it et al}, 
  {\it Detecting gravitational waves from cosmological phase transitions
  with LISA: an update,}
  JCAP {03} (2020) 024
  \arxiv{1910.13125}

\bibitem{rev_bb}
  D.~B\"odeker and W.~Buchm\"uller,
  {\it Baryogenesis from the weak scale to the grand unification scale,}
  Rev.\ Mod.\ Phys.\ {93} (2021) 035004
  \arxiv{2009.07294}

\bibitem{rev_mbh}
  M.B.~Hindmarsh, M.~L\"uben, J.~Lumma and M.~Pauly,
  {\it Phase transitions in the early universe,}
  SciPost Phys.\ Lect.\ Notes {24} (2021) 1
  \arxiv{2008.09136}

\bibitem{endpoint} 
  K.~Kajantie, M.~Laine, K.~Rummukainen and M.E.~Shaposhnikov,
  {\it Is there a hot electroweak phase transition
  at $\mH^{ } \gsim \mW^{ }$?,} 
  Phys.\ Rev.\ Lett.\ {77} (1996) 2887
  \arxiv{hep-ph/9605288}


\bibitem{twostep1}
  D.~Land and E.D.~Carlson,
  {\it Two stage phase transition in two Higgs models,}
  Phys.\ Lett.\ B {292} (1992) 107
  \arxiv{hep-ph/9208227}

\bibitem{twostep2}
  D.~B\"odeker, P.~John, M.~Laine and M.G.~Schmidt,
  {\it The 2-loop MSSM finite temperature effective potential
  with stop condensation,}
  Nucl.\ Phys.\ B {497} (1997) 387
  \arxiv{hep-ph/9612364}

\bibitem{twostep3}
  J.M.~Cline, G.D.~Moore and G.~Servant,
  {\it Was the electroweak phase transition preceded
  by a color broken phase?,}
  Phys.\ Rev.\ D {60} (1999) 105035
  \arxiv{hep-ph/9902220}

\bibitem{twostep4}
  S.~Profumo, M.J.~Ramsey-Musolf and G.~Shaughnessy,
  {\it Singlet Higgs phenomenology and the electroweak phase transition,}
  JHEP {08} (2007) 010
  \arxiv{0705.2425}

\bibitem{twostep5}
  J.R.~Espinosa, T.~Konstandin and F.~Riva,
  {\it Strong electroweak phase transitions in the Standard Model
  with a singlet,}
  Nucl.\ Phys.\ B {854} (2012) 592
  \arxiv{1107.5441}


\bibitem{landau5}
  L.D.~Landau and E.M.~Lifshitz,
  {\it Statistical Physics, Part 1}, \S162
  (Butterworth-Heinemann, Oxford).

\bibitem{eikr}
  K.~Enqvist, J.~Ignatius, K.~Kajantie and K.~Rummukainen,
  {\it Nucleation and bubble growth in a first-order
  cosmological electroweak phase transition,}
  Phys.\ Rev.\ D {45} (1992) 3415.

\bibitem{landau6}
  L.D.~Landau and E.M.~Lifshitz,
  {\it Fluid Mechanics}, \S132
  (Butterworth-Heinemann, Oxford).

\bibitem{ikkl}
  J.~Ignatius, K.~Kajantie, H.~Kurki-Suonio and M.~Laine,
  {\it Growth of bubbles in cosmological phase transitions,}
  Phys.\ Rev.\ D {49} (1994) 3854
  \arxiv{astro-ph/9309059}

\bibitem{simu}
  M.~Hindmarsh, S.J.~Huber, K.~Rummukainen and D.J.~Weir,
  {\it Shape of the acoustic gravitational wave power spectrum
  from a first order phase transition,}
  Phys.\ Rev.\ D {96} (2017) 103520; 
  {\it ibid.}\ {101} (2020) 089902 (erratum) 
  \arxiv{1704.05871}

\bibitem{turb}
  J.~Dahl, M.~Hindmarsh, K.~Rummukainen and D.J.~Weir,
  {\it Primordial acoustic turbulence: Three-dimensional
  simulations and gravitational wave predictions,}
  Phys.\ Rev.\ D {110} (2024) 103512
  \arxiv{2407.05826}

\bibitem{vw299} 
  H.~Kurki-Suonio and M.~Laine,
  {\it Real-Time History of the Cosmological Electroweak Phase Transition,}
  Phys.\ Rev.\ Lett.\ {77} (1996) 3951
  \arxiv{hep-ph/9607382}

\bibitem{defl}
  H.~Kurki-Suonio,
  {\it Deflagration bubbles in the quark-hadron phase transition,}
  Nucl.\ Phys.\ B {255} (1985) 231.

\bibitem{deto}
  M.~Laine,
  {\it Bubble growth as a detonation,}
  Phys.\ Rev.\ D {49} (1994) 3847
  \arxiv{hep-ph/9309242}

\bibitem{hks}
  H.~Kurki-Suonio and M.~Laine,
  {\it Supersonic deflagrations in cosmological phase transitions,}
  Phys.\ Rev.\ D {51} (1995) 5431
  \arxiv{hep-ph/9501216}


\bibitem{vw2}   %
  D.~B\"odeker and G.D.~Moore,
  {\it Electroweak bubble wall speed limit,}
  JCAP {05} (2017) 025
  \arxiv{1703.08215}

\bibitem{vw3}   %
  S.~H\"oche, J.~Kozaczuk, A.J.~Long, J.~Turner and Y.~Wang,
  {\it Towards an all-orders calculation of
  the electroweak bubble wall velocity,}
  JCAP {03} (2021) 009
  \arxiv{2007.10343}

\bibitem{vw45}  %
  A.~Azatov and M.~Vanvlasselaer,
  {\it Bubble wall velocity: heavy physics effects,}
  JCAP {01} (2021) 058
  \arxiv{2010.02590}

\bibitem{vw5}   %
  X.~Wang, F.P.~Huang and X.~Zhang,
  {\it Bubble wall velocity beyond leading-log approximation
  in electroweak phase transition,}
  \arxiv{2011.12903}

\bibitem{vw8}   %
  Y.~Gouttenoire, R.~Jinno and F.~Sala,
  {\it Friction pressure on relativistic bubble walls,}
  JHEP {05} (2022) 004
  \arxiv{2112.07686}

\bibitem{vw8aa}  %
  B.~Laurent and J.M.~Cline,
  {\it First principles determination of bubble wall velocity,}
  Phys.\ Rev.\ D {106} (2022) 023501
  \arxiv{2204.13120}

\bibitem{vw8a}
  W.-Y.\ Ai, 
  {\it Logarithmic divergent friction on ultrarelativistic bubble walls,}
  JCAP {10} (2023) 052
  \arxiv{2308.10679}

\bibitem{vw8b}
  A.~Azatov, G.~Barni, R.~Petrossian-Byrne and M.~Vanvlasselaer,
  {\it Quantisation across bubble walls and friction,}
  JHEP {05} (2024) 294
  \arxiv{2310.06972}

\bibitem{vw8c}
  S.~De Curtis, L.~Delle Rose, A.~Guiggiani, {\'A}.~Gil Muyor and G.~Panico,
  {\it Non-linearities in cosmological bubble wall dynamics,}
  JHEP {05} (2024) 009
  \arxiv{2401.13522}

\bibitem{vw8d}
  A.J.~Long and J.~Turner,
  {\it Thermal pressure on ultrarelativistic bubbles from
  a semiclassical formalism,}
  JCAP {11} (2024) 024
  \arxiv{2407.18196}

\bibitem{vw8e}
  A.~Ekstedt, O.~Gould, J.~Hirvonen, B.~Laurent, L.~Niemi,
  P.~Schicho and J.~van de Vis,
  {\it How fast does the WallGo? A package for computing
  wall velocities in first-order phase transitions,}
  JHEP {04} (2025) 101
  \arxiv{2411.04970}

\bibitem{vw8f}
  M.J.~Ramsey-Musolf and J.~Zhu,
  {\it Bubble wall velocity from Kadanoff-Baym equations:
  fluid dynamics and microscopic interactions,}
  \arxiv{2504.13724}

\bibitem{vw8g}
  W.-Y.~Ai, M.~Carosi, B.~Garbrecht, C.~Tamarit and M.~Vanvlasselaer,
  {\it Bubble wall dynamics from nonequilibrium quantum field theory,}
  JHEP {08} (2025) 077
  \arxiv{2504.13725}


\bibitem{vw30}  
  T.~Konstandin and J.M.~No,
  {\it Hydrodynamic obstruction to bubble expansion,}
  JCAP {02} (2011) 008
  \arxiv{1011.3735}

\bibitem{vw31}  
  M.~Barroso Mancha, T.~Prokopec and B.~Swiezewska,
  {\it Field-theoretic derivation of bubble-wall force,}
  JHEP {01} (2021) 070
  \arxiv{2005.10875}

\bibitem{vw32}  
  S.~Balaji, M.~Spannowsky and C.~Tamarit,
  {\it Cosmological bubble friction in local equilibrium,}
  JCAP {03} (2021) 051
  \arxiv{2010.08013}

\bibitem{vw33}   
  W.-Y.~Ai, B.~Garbrecht and C.~Tamarit,
  {\it Bubble wall velocities in local equilibrium,}
  JCAP {03} (2022) 015
  \arxiv{2109.13710}

\bibitem{vw34}  
  S.J.~Wang and Z.Y.~Yuwen,
  {\it Hydrodynamic backreaction force of cosmological bubble expansion,}
  Phys.\ Rev.\ D {107} (2023) 023501
  \arxiv{2205.02492}

\bibitem{vw36}  
  W.-Y.~Ai, B.~Laurent and J.~van de Vis,
  {\it Model-independent bubble wall velocities in local thermal equilibrium,}
  JCAP {07} (2023) 002
  \arxiv{2303.10171}
  
\bibitem{vw37}  
  T.~Krajewski, M.~Lewicki and M.~Zych,
  {\it Hydrodynamical constraints on the bubble wall velocity,}
  Phys.\ Rev.\ D {108} (2023) 103523
  \arxiv{2303.18216}

\bibitem{vw37a}
  W.-Y.~Ai, X.~Nagels and M.~Vanvlasselaer,
  {\it Criterion for ultra-fast bubble walls:
  the impact of hydrodynamic obstruction,}
  JCAP {03} (2024) 037
  \arxiv{2401.05911}

\bibitem{vw38}
  W.-Y.~Ai, B.~Laurent and J.~van de Vis,
  {\it Bounds on the bubble wall velocity,}
  JHEP {02} (2025) 119
  \arxiv{2411.13641}

\bibitem{vw39}
  T.~Krajewski, M.~Lewicki, I.~Na\l{}{e}cz and M.~Zych,
  {\it Steady-state bubbles beyond local thermal equilibrium,}
  JHEP {06} (2025) 118
  \arxiv{2411.16580}

\bibitem{vw41}
  M.~Carena, A.~Ireland, T.~Ou and I.R.~Wang,
  {\it The Discriminant Power of Bubble Wall Velocities:
  Gravitational Waves and Electroweak Baryogenesis,}
  \arxiv{2504.17841}

\bibitem{vw42}
  C.~Branchina, A.~Conaci, S.~De Curtis and L.~Delle Rose,
  {\it Electroweak Phase Transition and Bubble Wall Velocity
  in Local Thermal Equilibrium,}
  \arxiv{2504.21213}

\bibitem{vw43}
  Z.~Si, H.~Wang, L.~Wang, Y.~Xiao and Y.~Zhang,
  {\it The Bubble Wall Velocity in Local Thermal Equilibrium
  with Full Effective Potential,}
  \arxiv{2505.19584}
    

\bibitem{landau9} 
  E.M.~Lifshitz and L.P.~Pitaevskii, 
  {\it Statistical Physics, Part 2}, \S88-89
  (Butterworth-Heinemann, Oxford).

\bibitem{hydro}
  G.~Jackson and M.~Laine,
  {\it Hydrodynamic fluctuations from a weakly coupled scalar field,}
  Eur.\ Phys.\ J.\ C {78} (2018) 304
  \arxiv{1803.01871}


\bibitem{shear}
  P.~Kovtun, G.D.~Moore and P.~Romatschke,
  {\it Stickiness of sound: An absolute lower limit on viscosity
  and the breakdown of second-order relativistic hydrodynamics,}
  Phys.\ Rev.\ D {84} (2011) 025006
  \arxiv{1104.1586}


\bibitem{hk}
  P.Y.~Huet, K.~Kajantie, R.G.~Leigh, B.H.~Liu and L.D.~McLerran,
  {\it Hydrodynamic stability analysis of burning bubbles
  in electroweak theory and in QCD,}
  Phys.\ Rev.\ D {48} (1993) 2477
  \arxiv{hep-ph/9212224}


\bibitem{ae}
  A.~Ekstedt,
  {\it Bubble nucleation to all orders,}
  JHEP {08} (2022) 115
  \arxiv{2201.07331}

\bibitem{ae2}
  A.~Dashko and A.~Ekstedt,
  {\it Bubble-wall speed with loop corrections,}
  JHEP 03 (2025) 024
  \arxiv{2411.05075}

\bibitem{alica}
  M.~Laine, S.~Procacci and A.~Rogelj,
  {\it Evolution of coupled scalar perturbations through
   smooth reheating. Part I. Dissipative regime,}
  JCAP {10} (2024) 040
  \arxiv{2407.17074}


\bibitem{meg}
  M.E.~Carrington,
  {\it Effective potential at finite temperature in the Standard Model,}
  Phys.\ Rev.\ D {45} (1992) 2933.


\bibitem{pba}
  P.B.~Arnold,
  {\it Phase transition temperatures at next-to-leading order,}
  Phys.\ Rev.\ D {46} (1992) 2628
  \arxiv{hep-ph/9204228}

\bibitem{linde}
  A.D.~Linde,
  {\it Infrared problem in thermodynamics of the Yang-Mills gas,}
  Phys.\ Lett.\ {B 96} (1980) 289.



\bibitem{gv2}
  A.~Gynther and M.~Veps\"al\"ainen,
  {\it Pressure of the Standard Model near the electroweak phase transition,}
  JHEP {03} (2006) 011
  \arxiv{hep-ph/0512177}

\bibitem{lm}
  M.~Laine and M.~Meyer,
  {\it Standard Model thermodynamics across the electroweak crossover,}
  JCAP {07} (2015) 035
  \arxiv{1503.04935}

\bibitem{gt}
  O.~Gould and T.V.I.~Tenkanen,
  {\it Perturbative effective field theory expansions
  for cosmological phase transitions,}
  JHEP {01} (2024) 048
  \arxiv{2309.01672}


\bibitem{htl_old}
  T.S.~Bir\'o and M.H.~Thoma,
  {\it Damping rate and Lyapunov exponent of a Higgs field at high
  temperature,}
  Phys.\ Rev.\ D {54} (1996) 3465
  \arxiv{hep-ph/9603339}

\bibitem{higgswidth}
  M.~Eriksson and M.~Laine,
  {\it Soft contributions to the thermal Higgs width across
  an electroweak phase transition,}
  JCAP {06} (2024) 016
  \arxiv{2404.06116}


\bibitem{scan}
  P.~Klose, M.~Laine and S.~Procacci,
  {\it Gravitational wave background from vacuum and
  thermal fluctuations during axion-like inflation,}
  JCAP {12} (2022) 020
  \arxiv{2210.11710}


\bibitem{vw1}
  D.~B\"odeker and G.D.~Moore,
  {\it Can electroweak bubble walls run away?,}
  JCAP {05} (2009) 009
  \arxiv{0903.4099}
  
\end{thebibliography}
\end{document}